\documentclass[lettersize,journal]{IEEEtran}
\usepackage{amsmath,amsfonts}
\usepackage{array}
\usepackage{textcomp}
\usepackage{stfloats}
\usepackage{cite}
\usepackage{amsmath,amssymb,amsfonts}
\usepackage{bm}
\usepackage{graphicx}
\usepackage{textcomp}
\usepackage{xcolor}
\usepackage{caption}
\usepackage{makecell}
\usepackage[caption=false,font=footnotesize]{subfig}
\usepackage{balance}
\usepackage{adjustbox}

\usepackage[hyphens]{url}
\usepackage[breaklinks,colorlinks,linkcolor=black,citecolor=black,urlcolor=black]{hyperref}
\usepackage[hyphenbreaks]{breakurl}

\usepackage{algorithm,algpseudocode,setspace}

\makeatletter
\newcounter{phase}[algorithm]
\newlength{\phaserulewidth}

\makeatother

\begin{document}

\title{Meta-Learning-Based Fronthaul Compression for Cloud Radio Access Networks}

\author{\IEEEauthorblockN{Ruihua Qiao, Tao Jiang,~\IEEEmembership{Graduate Student Member,~IEEE}, and Wei Yu,~\IEEEmembership{Fellow,~IEEE}}
        % <-this % stops a space
\thanks{
The authors are with The Edward S. Rogers Sr. Department of Electrical and Computer Engineering, University of Toronto, Canada.
E-mails: \{ruihua.qiao, taoca.jiang\}@mail.utoronto.ca, weiyu@ece.utoronto.ca.
This work is supported by the Canada Research Chairs program and by Huawei Technologies Canada.
This work has been presented in part at the IEEE International Conference on Communications (ICC), 2023\cite{ruihua2023}.
Codes in this paper are available at: \url{https://github.com/RuihuaQiao/Meta-Learning-Fronthaul-Compression-CRAN}.
}}% <-this % stops a space



\maketitle

\begin{abstract}
This paper investigates the fronthaul compression problem in a user-centric cloud radio access network, in which single-antenna users are served by a central processor (CP) cooperatively via a cluster of remote radio heads (RRHs). 
To satisfy the fronthaul capacity constraint, this paper proposes a transform-compress-forward scheme, which consists of well-designed transformation matrices and uniform quantizers.
The transformation matrices perform dimension reduction in the uplink and dimension expansion in the downlink. 
To reduce the communication overhead for designing the transformation matrices, this paper further proposes a deep learning framework to first learn a suboptimal transformation matrix at each RRH based on the local channel state information (CSI), and then to refine it iteratively. 
To facilitate the refinement process, we propose an efficient signaling scheme that only requires the transmission of low-dimensional effective CSI and its gradient between the CP and RRH, and further, a meta-learning based gated recurrent unit network to reduce the number of signaling transmission rounds. 
For the sum-rate maximization problem, simulation results show that the proposed two-stage neural network can perform close to the fully cooperative global CSI based benchmark with significantly reduced communication overhead for both the uplink and the downlink.
Moreover, using the first stage alone can already outperform the existing local CSI based benchmark.    
\end{abstract}


\section{Introduction}

Cloud radio access network (C-RAN) \cite{checko2014cloud,simeone2016cloud,peng2015fronthaul}, also known as cell-free multiple-input
multiple-output (MIMO) network\cite{9064545,elhoushy2021cell,masoumi2019performance}, is envisioned as a key building block for future wireless networks. 
In a C-RAN system, the uplink or downlink signals between the users and the remote radio heads (RRHs) are jointly encoded or decoded in the central processor (CP), thereby effectively addressing the issue of inter-cell interference \cite{yu2017cooperative}.
However, due to the limited fronthaul capacity, it is necessary for each RRH to compress the uplink received signal vectors or for the CP to compress the downlink beamformed signals before forwarding them onto the fronthaul.
In the uplink, the compression can be done by applying a dimension-reducing transformation matrix followed by a scalar quantizer to the received signals at each RRH\cite{liu2015optimized, liu20171, wiffen2019distributed, wiffen2020mf,wiffen2021distributed}.
In the downlink, the compression can be achieved by applying a scalar quantizer to the beamformed signals in the CP followed by a dimension-expansion transformation matrix at each RRH\cite{arad2018precode,wiffen2021distributed}.
We refer to this scheme as the \textit{transform-compress-forward} strategy.
The objective of this paper is to design a set of transformation matrices so that simple scalar quantizers can achieve high compression efficiency.
This is a challenging problem because the optimal transformation matrices need to be designed jointly at the CP using global channel state information (CSI) \cite{wiffen2020mf,wiffen2021distributed,schizas2007distributed}, which requires significant communication overhead for sharing the high-dimensional CSI and the designed transformation matrices among the RRHs and the CP.
To address this issue, this paper demonstrates that by embracing a machine learning approach, particularly employing a specifically designed gated recurrent unit (GRU) meta-learning framework, we can find near-optimal transformation matrices with very small communication overhead.

As already mentioned, the joint design of the optimal transformation matrices across the RRHs requires global CSI.
To reduce the communication overhead for CSI sharing, heuristic methods using local CSI, e.g., eigenvalue decomposition (EVD) based method, are proposed in \cite{liu20171,liu2015optimized,wiffen2019distributed,arad2018precode}, but its performance is highly suboptimal since it minimizes the local reconstruction error at each RRH without accounting for the end-to-end system objective.
The work \cite{sohrabi2022learning} proposes a data-driven approach to optimize the end-to-end mean squared error (MSE) based on local CSI, which outperforms the EVD based method due to the use of the end-to-end objective, but there is still a large gap to the global CSI based benchmark.

The goal of this paper is to design near-optimal transformation matrices with as small communication overhead of exchanging CSI as possible.
Toward this end, this paper proposes a novel two-stage deep learning framework.
In the first stage, transformation matrices are derived from the local CSI at the RRHs using fully connected deep neural networks (DNNs).
In the second stage, the designed transformation matrices are further refined iteratively using the gradient of the system objective with respect to the transformation matrices.
To reduce the signaling dimension for gradient transmission in each iteration, this paper proposes an efficient signaling strategy, which allows the gradient at each RRH to be calculated from a low-dimensional signal from the CP.
Further, to reduce the overall refinement rounds, this paper proposes a meta-learning based GRU network that can learn an efficient refinement update based on historical and current gradients.
Simulation results show that the trained neural network with only the first stage can already outperform the existing local CSI based benchmark without introducing additional communication costs.
Moreover, when a few iterations are allowed in the second stage, the proposed network can achieve almost the same performance as the global CSI based benchmark but with a significant reduction in communication overhead.

\subsection{Prior Works}
\subsubsection{Uplink Fronthaul Compression}
The uplink C-RAN compression schemes are mostly based on the \textit{compress-forward} strategy, where the received user signals at the RRHs are quantized into digital codewords, entropy encoded, and finally forwarded onto the digital fronthaul.
The design of optimal quantizers is investigated in \cite{zhou2016optimal} and references therein, where Wyner-Ziv coding and vector quantization are proposed to take advantage of the correlation of the received signal across different RRHs and across different antennas of each RRH.
However, the Wyner-Ziv scheme and vector quantizers are computationally costly to implement, which restricts their use in practice. 
Some low-complexity approaches that use uniform scalar quantizers are proposed under different system setups.
For example, \cite{liu2015joint} considers an orthogonal frequency-division multiple access system and proposes to use bits allocation on top of the scalar quantizer of each subcarrier through bisection search.
The work \cite{liu20171} considers a narrowband communication scenario and assigns equal bits to the scalar quantizer corresponding to different dimensions of the received signal at each RRH.
However, these approaches tend to be suboptimal in terms of performance.

The generalization of the compress-forward strategy, termed in this paper as the \textit{transform-compress-forward} strategy, has also been proposed in previous literature under the name of beamform-compress-forward or spatial compression.
Compared to the compress-forward scheme, it adds a linear dimension reduction step before quantization, which can result in a significant performance gain over simple uniform quantization when the RRHs are equipped with a large number of antennas but only schedules a small number of users \cite{liu20171}.
This compression strategy can be regarded as a transform coding scheme from the perspective of continuous source coding in information theory \cite{goyal2001theoretical}.
The intuition of using the dimension-reducing matrices is that restricting the high-dimensional signals to a suitable subspace can improve the efficiency of the uniform scalar quantizer afterwards.

However, finding the optimal dimension-reducing matrices is a highly nontrivial task.
The main challenge is due to the fact that the system objective (e.g., the system sum rate) is a function of all transformation matrices, so the optimal transformation matrices should be jointly designed in the CP using the CSI from all the RRHs.
Obtaining such global CSI requires significant communication overhead between the RRHs and the CP.
Moreover, assuming the availability of global CSI at the CP,  the resulting optimization problem is also highly nonconvex.
The work \cite{wiffen2020mf} proposes a heuristic channel selection based matched filtering scheme to maximize the joint mutual information between all compressed signals and original full-dimensional signals.
The works \cite{wiffen2021distributed,schizas2007distributed} propose block coordinate descent based algorithms to design transformation matrices under different objectives. 

To reduce the communication overhead for CSI transmission, local CSI based methods are proposed in \cite{liu20171,liu2015optimized,wiffen2019distributed, sohrabi2022learning,wiffen2021distributed}, where transformation matrices are designed at each RRH individually using the available local CSI.
Specifically, \cite{liu20171,liu2015optimized,wiffen2019distributed} propose to design each transformation matrix using the eigenvectors corresponding to the largest eigenvalues of the covariance matrix of the received signal at each RRH. 
However, this EVD based approach can be highly suboptimal since it minimizes the local reconstruction error at each RRH instead of the end-to-end system objective.
An improved version of EVD is proposed in \cite{wiffen2021distributed}, where the global large-scale fading statistics are assumed to be available at each RRH and are incorporated into the local covariance matrices.
Moreover, \cite{sohrabi2022learning} addresses a similar distributed estimation problem in a data-driven manner, where local DNNs that map the local CSI into transformation matrices are trained to optimize the end-to-end MSE.
However, although the two approaches in \cite{wiffen2021distributed} and \cite{sohrabi2022learning} outperform the EVD based method due to the availability of the global statistical information and the use of an end-to-end objective function, their performance still leaves a large gap to the global CSI based method.

\subsubsection{Downlink Fronthaul Compression}

In the C-RAN downlink, the user symbols can be jointly beamformed and further compressed in the CP before being sent to the RRHs through the downlink fronthaul channel.
Similar to the uplink, the \textit{compress-forward} and \textit{transform-compress-forward} strategies can also be applied to the downlink.
Specifically, the compress-forward scheme directly quantizes the beamformed signals\cite{patil2018hybrid,park2013joint, liu20171}, while
the transform-compress-forward scheme includes an additional dimension expansion step\cite{arad2018precode,wiffen2021distributed}.
This paper focuses on the transform-compress-forward scheme in the downlink due to its better performance.
Specifically, the CP first computes low-dimensional beamformed signals, which are then quantized and transmitted to the RRHs through the downlink fronthaul.
Then, the RRHs apply a transformation matrix to the low-dimensional quantized signals to increase the dimension of the transmitted fronthaul signal to be equal to that of the RRH antennas. 
Similar to the uplink, the key is to design the appropriate dimension-expansion transformation matrices at the RRHs.
A simple heuristic that selects the strongest users for each RRH using local CSI is proposed in \cite{arad2018precode}.
Further, the work \cite{wiffen2021distributed} suggests the feasibility of using the uplink dimension-reducing matrices as the downlink dimension-expansion matrices at the RRHs, which can be designed using either global CSI or local CSI.
However, the same CSI communication overhead problem exists for the downlink as in the uplink.
This paper aims to design a unified scheme applicable for both uplink and downlink that can efficiently trade off between communication overhead and the system sum rate performance.

It is worth noting that the \textit{data-sharing} approach proposed in \cite{liu2016joint, dai2014sparse, liu20171} is also a potential candidate in the C-RAN downlink. 
In the data-sharing approach, the CP first sends the RRHs the designed beamforming coefficients at the beginning of each channel coherence block.
Then, in the data transmission stage, the RRHs receive raw user data from the CP, perform the beamforming operation, and then transmit the beamformed signal to the users.
However, the work \cite{patil2018hybrid} shows that its performance is inferior when compared to the compress-forward strategy except for extremely low fronthaul capacity scenarios where the quantization error of the compression strategy becomes too large. 
This is mainly because the user symbol needs to be repeatedly transmitted to all the RRHs within its cluster, which severely limits the cooperation cluster size in systems with limited fronthaul.

\subsection{Main Contributions}
The main contributions of this paper are summarized as follows:
\begin{enumerate}
    \item This paper proposes a general framework that can solve both the uplink and downlink fronthaul compression problems. 
    Specifically, in the uplink, the proposed algorithm first designs the transformation matrices at the RRHs that are used to reduce the dimension of the received uplink user signals, then applies scalar quantizers to quantize the dimension-reduced signals.
    Analogously, in the downlink, the proposed algorithm applies scalar quantizers at the CP to quantize the low-dimensional beamformed signal vectors to be transmitted to the RRHs, 
    then designs the dimension-expansion matrices at the RRHs to increase the dimension of the beamformed signals to the dimension of the RRH antennas.
     
    \item  The proposed framework can design transformation matrices with smaller communication overhead without much loss of performance as compared to the conventional schemes using global CSI.
    The proposed framework is data-driven and includes two stages.
    In the first stage, initial transformation matrices are mapped from local CSI using local DNNs that are jointly trained using the objective of the end-to-end system sum rate.
    Compared to previous local CSI based methods that rely on the heuristic of minimizing the local reconstruction error at each individual RRH, the proposed data-driven scheme uses the global objective function in the training phase, so it enables the local DNNs at the RRHs to learn to cooperate implicitly to some extent even if the exact realization of global CSI is not available.
    In the second stage, this paper proposes a communication-efficient scheme to trade off communication overhead with the system sum rate performance.
    Specifically, instead of transmitting the high-dimensional global CSI, a low-dimensional signaling scheme is proposed to allow the RRHs and the CP to communicate global information with small communication overhead in each iteration.
    Moreover, to reduce the number of overall iteration rounds, this paper proposes a novel GRU based meta-learning method to accelerate the convergence speed of the iterative algorithm.

    \item Simulation results show that the proposed two-stage algorithm can be applied to both the uplink and the downlink to achieve excellent performance under a practical simulation setup that involves user-centric clustering and a 19-cell wrap-around layout.
    Specifically, using the first stage alone can already outperform the existing local CSI based heuristic algorithms. 
    Moreover, the system objective function converges quickly as the number of communication rounds increases in the meta-learning based GRU network in the second stage, allowing us to achieve almost the same performance as the global CSI based benchmark with significantly reduced communication overhead.

\end{enumerate}

\subsection{Paper Organization and Notations}
The remainder of this paper is organized as follows.
Section \ref{sec:model} presents the uplink and downlink C-RAN system models, and unifies them to formulate the problem of designing the transformation matrices under the transform-compress-forward strategy.
Section \ref{sec:design_W} introduces the proposed two-stage data-driven method for designing the transformation matrices at the RRHs.
Section \ref{sec:quantization} describes the uniform quantization scheme used to compress the fronthaul signals.
As two applications of the proposed algorithm, Section \ref{sec:ul} and \ref{sec:dl} formulate the uplink and downlink sum-rate maximization problems, respectively, and present numerical results to validate the effectiveness of the proposed scheme.
Finally, Section \ref{sec:conclusion} concludes the paper.

This paper adopts the notation that lower-case letters represent scalar variables, lower-case bold-face letters denote column vectors, and upper-case bold-face letters denote matrices. The superscripts $(\cdot)^{\mathsf{H}}$ and $(\cdot)^{-1}$ denote the Hermitian transpose and the inverse operations, respectively. The identity matrix with appropriate dimensions is represented by $\mathbf{I}$; $\mathbb{C}^{m \times n}$ represents an $m$ by $n$ dimensional complex space; $\mathcal{N}(\mathbf{0}, \bm{\Sigma})$ denotes the zero-mean circularly symmetric complex Gaussian distribution with covariance matrix $\bm{\Sigma}$; and $\mathcal{U}(a, b)$ represents a uniform distribution on the interval $[a, b]$. 
The notations $\log (\cdot)$ and $\mathbb{E}[\cdot]$ represent the binary logarithm and expectation operators, respectively.
The function ${\mathrm{diag}}\left( \cdot \right)$ represents a diagonal matrix with the arguments as its diagonal elements.
Additionally, $\|\bm{v}\|_2$ indicates the Euclidean norm of a vector $\bm{v}$, $|{v}|_2$ denotes the absolute value of the complex scalar $v$, and $|{S}|$ represents the cardinality of a set ${S}$. Finally, the hyperbolic tangent activation function is defined as $\tanh (x) \triangleq \frac{e^x-e^{-x}}{e^x+e^{-x}}$.

\section{System Model and Problem Formulation}\label{sec:model}

Consider a C-RAN system where $N$ single antenna users are served by the CP through $B$ geographically distributed RRHs, each equipped with $M$ antennas. 
The fronthaul channels between the CP and the RRHs are assumed to be digital links with finite capacities.
To mitigate the inter-cell interference, the encoding and decoding of the user messages are performed jointly at the CP.
If the fronthaul links between the RRHs and the CP had infinite capacity, the RRHs can simply forward the received signal to the CP in the uplink; likewise, the CP can simply forward the beamformed signals to the RRHs in the downlink. 
But in practical systems where the fronthaul links have finite capacity,
a compression step would be needed before forwarding the respective uplink and downlink signals onto the fronthaul.
Specifically, this paper adopts a transform coding approach, named \textit{transform-compress-forward}, which consists of a linear transformation step and a uniform quantization step.
In the following, we show that the problem of designing the transformation matrices at the RRHs in both the uplink and the downlink can be formulated in a unified fashion.

\begin{figure*}[t]
\centering
\includegraphics[width=0.7\linewidth]{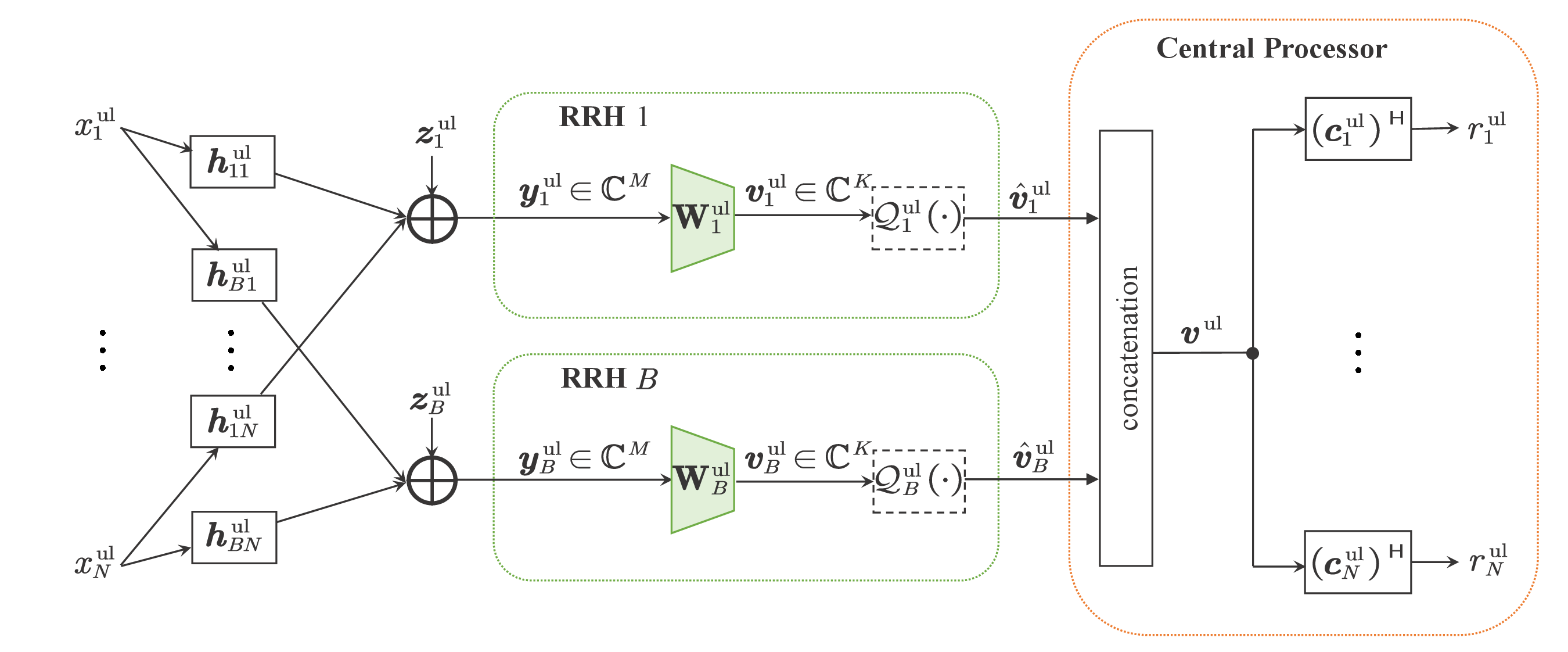}
\caption{Uplink C-RAN system model. The transformation matrix $\mathbf{W}_b^{\mathrm{ul}}$ at RRH $b$ is designed using the proposed two-stage DNN to maximize the system utility. The quantizer $\mathcal{Q}_b^{\mathrm{ul}}\left(\cdot\right)$ is ignored when designing $\mathbf{W}_b^{\mathrm{ul}}$ and added after $\mathbf{W}_b^{\mathrm{ul}}$ is designed.}
\label{fig:sys_model_ul}
\end{figure*}

\subsection{Uplink System Model}\label{sec:ul_sys_general}

First, the uplink compression strategy for the C-RAN is described as follows.
As shown in Fig. \ref{fig:sys_model_ul}, the uplink received signal at RRH $b$ from the users can be written as:
\begin{equation}
    \bm{y}_b^{\mathrm{ul}} =  \mathbf{H}_b^{\mathrm{ul}}  \bm{x}^{\mathrm{ul}}+ \bm{z}_b^{\mathrm{ul}}  ,
    \label{eq:y_b_ul}
\end{equation}
where $\mathbf{H}_b^{\mathrm{ul}}=\left[ \bm{h}_{b1}^{\mathrm{ul}}\cdots \bm{h}_{bN}^{\mathrm{ul}} \right] \in \mathbb{C} ^{M\times N}$ with $\bm{h}_{bn}^{\mathrm{ul}} \in \mathbb{C}^M$ denoting the channel between RRH $b$ and user $n$, 
$\bm{x}^{\mathrm{ul}}=\left[ x_1^{\mathrm{ul}}\cdots x_N^{\mathrm{ul}} \right]^{\mathsf{H}} \in \mathbb{C} ^{N}$ with ${x}_n^{\mathrm{ul}}\sim \mathcal{CN}\left( 0, p_n^{\mathrm{ul}} \right) $ denoting the transmitted signal of user $n$,
and $\bm{z}_b^{\mathrm{ul}}\in \mathbb{C}^M \sim \mathcal{CN}\left( 0, \sigma_{\mathrm{ul}}^2 \mathbf{I}\right) $ is the additive white Gaussian noise (AWGN) at RRH $b$.
At RRH $b$, a \textit{dimension-reducing} linear transformation $\mathbf{W}_b^{\mathrm{ul}} \in \mathbb{C}^{K \times M}$ is applied to the signal $\bm{y}_b^{\mathrm{ul}} \in \mathbb{C}^M$ 
to produce $\bm{v}_b^{\mathrm{ul}} \in \mathbb{C}^K$ with $K < M$ as:
\begin{equation}
    \bm{v}_b^{\mathrm{ul}} = \mathbf{W}_b^{\mathrm{ul}}\bm{y}_b^{\mathrm{ul}} = \mathbf{F}_b^{\mathrm{ul}}\bm{x}^{\mathrm{ul}} + \mathbf{W}_b^{\mathrm{ul}}\bm{z}_b^{\mathrm{ul}},
    \label{eq:v_b_ul}
\end{equation}
where $\mathbf{F}_b^{\mathrm{ul}} \in \mathbb{C}^{K \times N}$ is the effective channel between RRH $b$ and all the users, given by:
\begin{equation}
    \mathbf{F}_b^{\mathrm{ul}}=\mathbf{W}_b^{\mathrm{ul}}\mathbf{H}_b^{\mathrm{ul}} .
    \label{eq:F_b_ul}
\end{equation} 
Next, a uniform scalar quantizer $\mathcal{Q}_b^{\mathrm{ul}}\left(\cdot \right)$ compresses the dimension-reduced signal $\bm{v}_b^{\mathrm{ul}}$ into the digital signal $\hat{\bm{v}}_b^{\mathrm{ul}}$.
We ignore the quantization error for now, i.e., assume $\hat{\bm{v}}_b^{\mathrm{ul}}={\bm{v}}_b^{\mathrm{ul}}$, and explain the details of designing the quantizer in Section \ref{sec:quantization}.

To reap the cooperation gain in the C-RAN architecture, the CP designs a linear minimum mean squared error (MMSE) receive beamformer \cite{joham2005linear} $\bm{c}_n^{\mathrm{ul}} \in \mathbb{C}^{BK}$ for user $n$ using the  effective CSI across all the RRHs, denoted by $\mathbf{{F}}^{\mathrm{ul}}=\left[ \left( \mathbf{F}_{1}^{\mathrm{ul}}\right)^{\mathsf{H}}, \dots, \left( \mathbf{F}_{B}^{\mathrm{ul}}\right)^{\mathsf{H}}\right]^{\mathsf{H}}\in \mathbb{C} ^{BK\times N}$, as:
\begin{equation}
   \bm{c}_n^{\mathrm{ul}}=\left( \mathbf{{F}}^{\mathrm{ul}} 
   \mathbf{P}^{\mathrm{ul}} \left(\mathbf{{F}}^{\mathrm{ul}}\right)^{\mathsf{H}}+\sigma _{\mathrm{ul}}^{2}  \mathbf{W}^{\mathrm{ul}}\left( \mathbf{W}^{\mathrm{ul}} \right) ^{\mathsf{H}} \right) ^{-1}\bm{f}_{n}^{\mathrm{ul}}
   ,
   \label{eq:cn_ul_W}
\end{equation}
where $\mathbf{P}^{\mathrm{ul}} = \mathrm{diag}\left( p_1^{\mathrm{ul}},\cdots, p_N^{\mathrm{ul}}  \right) $ is a diagonal matrix with the user transmitting power on the diagonal axis,
$\mathbf{W}^{\mathrm{ul}} = {\mathrm{diag}}\left(\mathbf{W}_1^{\mathrm{ul}}, \cdots, \mathbf{W}_B^{\mathrm{ul}}  \right)$ is the overall block-diagonal transformation matrix across all the RRHs,  
and $\bm{f}_n^{\mathrm{ul}}$ is the $n$-th column of $\mathbf{{F}}^{\mathrm{ul}}$ representing the effective channel vector between user $n$ and all the RRHs. 
Thus, the signal after receive beamforming for user $n$ can be written as:
\begin{equation}
    \begin{aligned}
    {r}_n^{\mathrm{ul}} 
    &= \left(\bm{c}_n^{\mathrm{ul}}\right)^{\mathsf{H}}\bm{v}^{\mathrm{ul}} \\
    & =\left( \bm{c}_{n}^{\mathrm{ul}} \right) ^{\mathsf{H}}\bm{f}_{n}^{\mathrm{ul}}x_n+\sum_{i\ne n}{\left( \bm{c}_{n}^{\mathrm{ul}} \right) ^{\mathsf{H}}\bm{f}_{i}^{\mathrm{ul}}x_i}+\left( \bm{c}_{n}^{\mathrm{ul}} \right) ^{\mathsf{H}}\mathbf{W}^{\mathrm{ul}}\bm{z}^{\mathrm{ul}},
\end{aligned}
\end{equation}
where $\bm{v}^{\mathrm{ul}} = \left[ \left(\bm{v}_{1}^{\mathrm{ul}} \right)^{\mathsf{H}}, \cdots, \left(\bm{v}_{B}^{\mathrm{ul}} \right)^{\mathsf{H}} \right]^{\mathsf{H}} \in \mathbb{C}^{BK}$ is the concatenated transformed signal across all the RRHs, 
and $\bm{z}^{\mathrm{ul}} = \left[ \left(\bm{z}_{1}^{\mathrm{ul}} \right)^{\mathsf{H}}, \cdots, \left(\bm{z}_{B}^{\mathrm{ul}} \right)^{\mathsf{H}} \right]^{\mathsf{H}} \in \mathbb{C}^{BM}$ is the  noise vector.
Finally, the uplink achievable rate of user $n$ can be written as:
\begin{equation}
    R_{n}^{\mathrm{ul}}=\log \left( 1+\frac{p_{n}^{\mathrm{ul}}\left| \left( \bm{c}_{n}^{\mathrm{ul}} \right) ^{\mathsf{H}}\bm{f}_{n}^{\mathrm{ul}} \right|^2}{\sum\limits_{\substack{ i\ne n}}{p_{i}^{\mathrm{ul}}\left| \left( \bm{c}_{n}^{\mathrm{ul}} \right) ^{\mathsf{H}}\bm{f}_{i}^{\mathrm{ul}} \right|^2+\sigma _{\mathrm{ul}}^{2}\left\| \left( \mathbf{W}^{\mathrm{ul}} \right) ^{\mathsf{H}}\bm{c}_{n}^{\mathrm{ul}} \right\| ^2}} \right)
    .
\label{eq:rate_R_ul_simple_0}
\end{equation}

It can be observed that the user rate $R_n^{\mathrm{ul}}$ is a function of the transformation matrices $\left\{ \mathbf{W}_b^{\mathrm{ul}} \right\} _{b=1}^{B}$ through the beamforming vectors $\bm{c}_n^{\mathrm{ul}}$ and effective channels $\left\{ \bm{f}_n^{\mathrm{ul}} \right\} _{n=1}^{N}$.
For example, the numerator $p_n^{\mathrm{ul}} \left|\left( \bm{c}_{n}^{\mathrm{ul}} \right) ^{\mathsf{H}}\bm{f}_{n}^{\mathrm{ul}}\right|^2$ can be expanded as:
\begin{equation}
\begin{aligned}
    \left|\left( \bm{c}_{n}^{\mathrm{ul}} \right) ^{\mathsf{H}}\bm{f}_{n}^{\mathrm{ul}}\right|^2
    =
    &\left|\left( 
    \bm{h}_{n}^{\mathrm{ul}} \right) ^{\mathsf{H}}\left( \mathbf{W}^{\mathrm{ul}} \right) ^{\mathsf{H}} 
    \left( \mathbf{W}^{\mathrm{ul}}\mathbf{H}^{\mathrm{ul}}
    \mathbf{P}^{\mathrm{ul}} 
    \left( \mathbf{H}^{\mathrm{ul}} \right) ^{\mathsf{H}}\left( \mathbf{W}^{\mathrm{ul}} \right) ^{\mathsf{H}} \right.\right.
    \\
    & ~~ + \left.\left. \sigma _{\mathrm{ul}}^{2}\mathbf{W}^{\mathrm{ul}}\left( \mathbf{W}^{\mathrm{ul}} \right) ^{\mathsf{H}} \right) ^{-1}\mathbf{W}^{\mathrm{ul}}\bm{h}_{n}^{\mathrm{ul}}\right|^2 ,
\end{aligned}
\label{eq:R_ul_numerator}%
\end{equation}
where $\mathbf{{H}}^{\mathrm{ul}}=\left[ \left( \mathbf{H}_{1}^{\mathrm{ul}}\right)^{\mathsf{H}}, \cdots, \left( \mathbf{H}_{B}^{\mathrm{ul}}\right)^{\mathsf{H}} \right]^{\mathsf{H}}\in \mathbb{C} ^{BM\times N}$ is the channel matrix between all the RRHs and all the users.
Thus, we can formulate a problem of optimizing the transformation matrices $\left\{ \mathbf{W}_b^{\mathrm{ul}} \right\} _{b=1}^{B}$ so as to maximize a system utility objective, e.g., the sum rate $\sum_{n}{R_n^{\mathrm{ul}}}$. 
Intuitively, the choice of $\left\{ \mathbf{W}_b^{\mathrm{ul}} \right\} _{b=1}^{B}$ determines the appropriate low-dimensional subspace onto which the received signals at the RRHs can be compressed.

\begin{figure*}[t]
\centering
\includegraphics[width=0.7\linewidth]{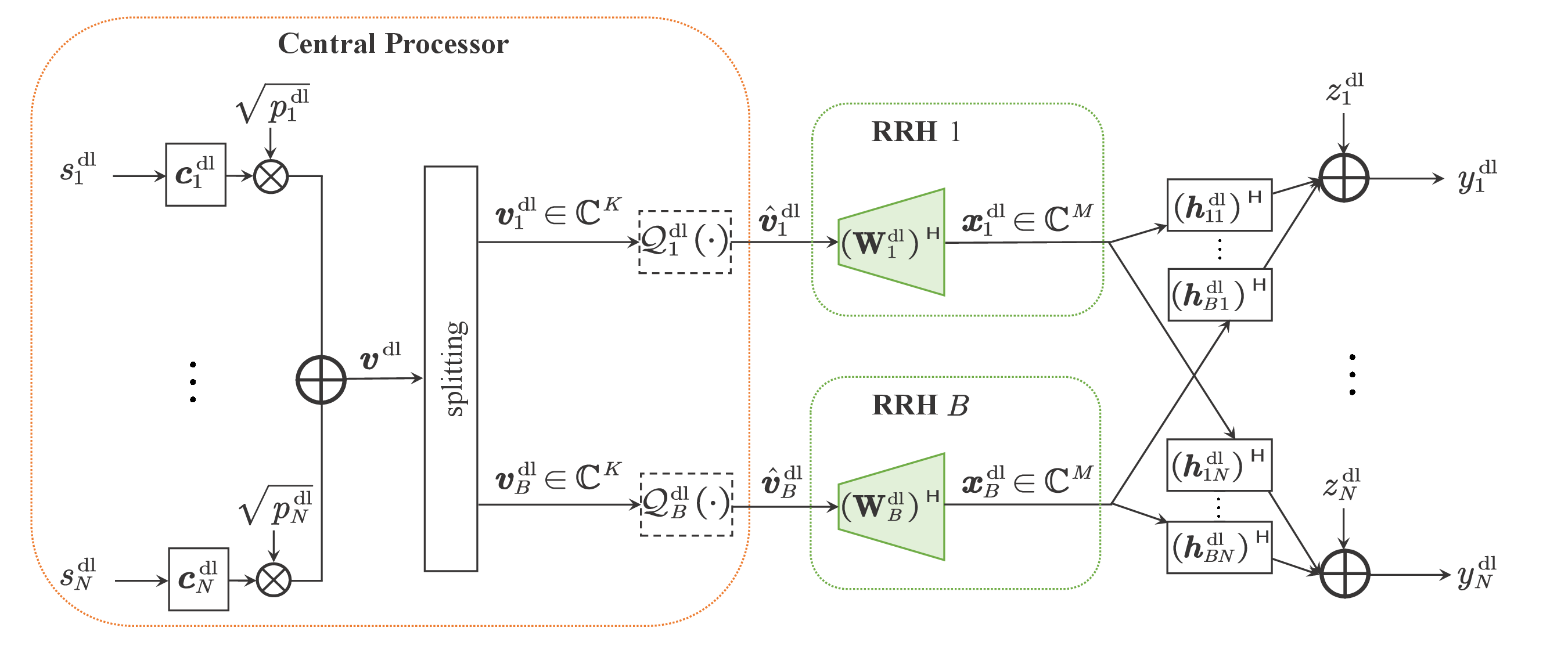}
\caption{Downlink C-RAN system model. The expansion matrix $\mathbf{W}_b^{\mathrm{dl}}$ at RRH $b$ is designed using the proposed two-stage DNN to maximize the system utility. The quantizer $\mathcal{Q}_b^{\mathrm{dl}}\left(\cdot\right)$ is ignored when designing $\mathbf{W}_b^{\mathrm{dl}}$ and added after $\mathbf{W}_b^{\mathrm{dl}}$ is designed.}
\label{fig:sys_model_dl}
\end{figure*}

Further, note that left-multiplying $\mathbf{W}^{\mathrm{ul}}$ by an invertible matrix $\mathbf{A}$, i.e., $\tilde{\mathbf{W}}^{\mathrm{ul}}=\mathbf{A} \mathbf{W}^{\mathrm{ul}}$, does not affect the maximum achievable rate in the uplink as the mutual information $\mathcal{I} \left( x_n^{\mathrm{ul}}; {r}_n^{\mathrm{ul}} \right)$ remains unchanged.
Specifically, we can observe from \eqref{eq:R_ul_numerator} that such an operation does not change the value of the numerator in \eqref{eq:rate_R_ul_simple_0};
it also keeps the interference power and noise power in the denominator in \eqref{eq:rate_R_ul_simple_0} unchanged.
This allows us to remove $\mathbf{W}_b^{\mathrm{ul}}\left(\mathbf{W}_{b}^{\mathrm{ul}}\right)^{\mathsf{H}}$ in the noise power term in \eqref{eq:rate_R_ul_simple_0} by choosing, without loss of generality, $\mathbf{W}_b^{\mathrm{ul}}$ to be a semi-orthogonal matrix, i.e., $\mathbf{W}_b^{\mathrm{ul}}\left(\mathbf{W}_{b}^{\mathrm{ul}}\right)^{\mathsf{H}}=\mathbf{I}$. 
This can be achieved by performing a QR decomposition on an unnormalized transformation matrix $\left(\tilde{\mathbf{W}}_b^{\mathrm{ul}}\right)^{\mathsf{H}}=\mathbf{Q} \mathbf{R}$ where $\mathbf{Q} \in \mathbb{C}^{M \times K}$ has orthonormal columns and $\mathbf{R} \in \mathbb{C}^{K\times K}$ is an upper triangular matrix,
and using the semi-orthogonal matrix $\mathbf{Q}^\mathsf{H}$ as $\mathbf{W}_b^{\mathrm{ul}}$.

With the semi-orthogonal $\mathbf{W}_b^{\mathrm{ul}}$ as chosen above, the rate in \eqref{eq:rate_R_ul_simple_0} can now be rewritten as:
\begin{equation}
    R_{n}^{\mathrm{ul}}=\log \left( 1+\frac{p_{n}^{\mathrm{ul}}\left| \left( \bm{c}_{n}^{\mathrm{ul}} \right) ^{\mathsf{H}}\bm{f}_{n}^{\mathrm{ul}} \right|^2}{\sum\limits_{\substack{ i\ne n}}{p_{i}^{\mathrm{ul}}\left| \left( \bm{c}_{n}^{\mathrm{ul}} \right) ^{\mathsf{H}}\bm{f}_{i}^{\mathrm{ul}} \right|^2+\sigma _{\mathrm{ul}}^{2}\left\| \bm{c}_{n}^{\mathrm{ul}} \right\| ^2}} \right) ,
\label{eq:rate_R_ul_simple}
\end{equation}
where $\bm{c}_n^{\mathrm{ul}}$ can be simplified from \eqref{eq:cn_ul_W} as:
\begin{equation}
    {\bm{c}}_n^{\mathrm{ul}}=\left( \mathbf{{F}}^{\mathrm{ul}} 
    \mathbf{P}^{\mathrm{ul}} \left(\mathbf{{F}}^{\mathrm{ul}}\right)^{\mathsf{H}}+\sigma _{\mathrm{ul}}^{2} \mathbf{I} \right) ^{-1}\bm{f}_{n}^{\mathrm{ul}} .
    \label{eq:c_n_ul_sec2}
\end{equation}

\subsection{Downlink System Model}

In the downlink, the CP communicates the data symbols $\left\{ 
s_n^{\mathrm{dl}} \right\}_{n=1}^N$ to the users via the RRHs.
In this paper, we explore a downlink beamforming strategy that treats the RRHs as distributed antennas.
Specifically, the CP pre-designs a beamformer for each user across all the RRHs.
In the data transmission phase, the CP calculates the beamformed signal vector and then sends the corresponding components to each of the RRHs.
Next, each RRH $b$ sends the beamformed signal component $\bm{x}_{b}^{\mathrm{dl}}\in \mathbb{C}^{M}$ to the users over the air.
The received signal of user $n$ now becomes:
\begin{equation}
    y_{n}^{\mathrm{dl}}=\sum_{b=1}^B{\left( \bm{h}_{bn}^{\mathrm{dl}} \right) ^{\mathsf{H}}\bm{x}_{b}^{\mathrm{dl}}}+z_{n}^{\mathrm{dl}} ,
    \label{eq:rn_dl}
\end{equation}
where $\bm{h}_{bn}^{\mathrm{dl}} \in \mathbb{C}^{M}$ denotes the channel vector between RRH $b$ and user $n$, 
$\bm{x}_b^{\mathrm{dl}}$ contains the user data symbols $\left\{ 
s_n^{\mathrm{dl}} \right\}_{n=1}^N$,
and ${z}_n^{\mathrm{dl}}\in \mathbb{C}\sim \mathcal{CN}\left( 0, \sigma_{\mathrm{dl}}^2 \right) $ denotes the AWGN at user $n$.
The system model in the downlink is shown in Fig. \ref{fig:sys_model_dl}.

Due to the limited fronthaul capacity, the CP can only send a compressed version of the beamformed signals to the RRHs.
This paper pursues a \textit{transform-compress-forward} strategy, in which the CP designs a low-dimensional signal ${\bm{v}}_b^{\mathrm{dl}} \in \mathbb{C}^K$ and transmits its quantized version $\hat{\bm{v}}_b^{\mathrm{dl}}$ to RRH $b$;
then, RRH $b$ uses a \textit{dimension-expansion} matrix $\mathbf{W}_b^{\mathrm{dl}} \in \mathbb{C}^{K \times M}$ with $M>K$ to expand the signal $\hat{\bm{v}}_b^{\mathrm{dl}}$ received from the CP to the signal ${\bm{x}}_b^{\mathrm{dl}} \in \mathbb{C}^{M}$ to be transmitted on its antennas.
The idea of the transformation step is that  ${\bm{v}}_b^{\mathrm{dl}}$ has the dimension of $K$ per RRH, instead of $M$, so it is easier to quantize at the CP.

Assuming that the transformation matrices $\left\{\mathbf{W}_b^{\mathrm{dl}}\right\}_{b=1}^B$ have been designed and fixed at the RRHs, the transmit beamformers $\left\{ \bm{c}_n^{\mathrm{dl}} \in \mathbb{C}^{BK}\right\}_{n=1}^N$ for all the users can be jointly designed in the CP by regarding the combined transformation matrices and the original channel matrices as low-dimensional effective channel matrices.
Specifically, the effective channel matrix $\mathbf{F}_b^{\mathrm{dl}}$ between RRH $b$ and all the users is given by:
\begin{equation}
    \mathbf{F}_b^{\mathrm{dl}} = \mathbf{W}_b^{\mathrm{dl}} \mathbf{H}_b^{\mathrm{dl}} ,
    \label{eq:F_b_dl}
\end{equation}
where $\mathbf{H}_b^{\mathrm{dl}}=\left[ \bm{h}_{b1}^{\mathrm{dl}}\cdots \bm{h}_{bN}^{\mathrm{dl}} \right] \in \mathbb{C} ^{M\times N}$ is the  channel matrix between RRH $b$ and all the users.
Now, the design of the optimal downlink transmit beamformer is, in general, a challenging task; many heuristic algorithms exist in literature, including matched filter, zero-forcing, regularized zero-forcing\cite{peel2005vector,choi2021large}, truncated polynomial expansion\cite{mueller2016linear}, and phased zero-forcing\cite{liang2014low}, etc.
This paper considers an approach based on the principle of uplink-downlink duality to design the downlink beamforming vector $\bm{c}_n^{\mathrm{dl}}$ for user $n$ using the transmit MMSE beamformer \cite{bjornson2014optimal}, as: 
\begin{equation}
    \bm{c}_{n}^{\mathrm{dl}}=\frac{\left( \frac{P^{\mathrm{dl}}}{N}\mathbf{F}^{\mathrm{dl}}\left( \mathbf{F}^{\mathrm{dl}} \right) ^{\mathsf{H}}+\sigma _{\mathrm{dl}}^{2}\mathbf{I} \right) ^{-1}\bm{f}_{n}^{\mathrm{dl}}}{\left\| \left( \frac{P^{\mathrm{dl}}}{N}\mathbf{F}^{\mathrm{dl}}\left( \mathbf{F}^{\mathrm{dl}} \right) ^{\mathsf{H}}+\sigma _{\mathrm{dl}}^{2}\mathbf{I} \right) ^{-1}\bm{f}_{n}^{\mathrm{dl}} \right\|}
     ,
    \label{eq:c_n_dl_sec2_0}
\end{equation}
where $P^{\mathrm{dl}}$ is the total downlink power budget across all the RRHs, 
$\mathbf{{F}}^{\mathrm{dl}}=\left[ \left( \mathbf{F}_{1}^{\mathrm{dl}}\right)^{\mathsf{H}}, \cdots, \left( \mathbf{F}_{B}^{\mathrm{dl}}\right)^{\mathsf{H}}\right]^{\mathsf{H}}\in \mathbb{C} ^{BK\times N}$ is the  effective channel matrix across all the RRHs,
and $\bm{f}_n^{\mathrm{dl}}$ is the $n$-th column of $\mathbf{{F}}^{\mathrm{dl}}$ representing the effective channel vector between user $n$ and all the RRHs.

Fixing the beamformers $\left\{ \bm{c}_n^{\mathrm{dl}} \right\}_{n=1}^N$, the combined beamformed signal $\bm{v}^{\mathrm{dl}} = \left[ \left(\bm{v}_{1}^{\mathrm{dl}} \right)^{\mathsf{H}}, \cdots, \left(\bm{v}_{B}^{\mathrm{dl}} \right)^{\mathsf{H}} \right]^{\mathsf{H}} \in \mathbb{C}^{BK}$, whose component $\bm{v}_b^{\mathrm{dl}}$ is intended for RRH $b$, can be computed as:
\begin{equation}
    \bm{v}^{\mathrm{dl}} = \sum_{n=1}^N {\sqrt{p_n^{\mathrm{dl}}} \bm{c}_n^{\mathrm{dl}} s_n^{\mathrm{dl}}} ,
\label{eq:v_dl}
\end{equation}
where ${s}_n^{\mathrm{dl}}\sim \mathcal{CN}\left( 0, 1 \right) $ is the user symbol,
and the power factors $\left\{p_n^{\mathrm{dl}}\right\}_{n=1}^N$ are chosen to satisfy the transmit power constraints at the RRHs.
 
Similar to the uplink fronthaul compression, a uniform scalar quantizer $\mathcal{Q}_b^{\mathrm{dl}}\left(\cdot \right)$  can be used to compress the continuous signal $\bm{v}_b^{\mathrm{dl}} \in \mathbb{C}^K$ into the digital signal $\hat{\bm{v}}_b^{\mathrm{dl}}$ at the CP, and transmit it to RRH $b$ through the fronthaul.
Then, RRH $b$ uses the transformation matrix $\mathbf{W}_b^{\mathrm{dl}} \in \mathbb{C}^{K \times M}$ to increase the dimension of the signal $\hat{\bm{v}}_b^{\mathrm{dl}} \in \mathbb{C}^{K}$ to $\bm{x}_b^{\mathrm{dl}} \in \mathbb{C}^{M}$ with $M> K$ as $\bm{x}_b^{\mathrm{dl}} = \left(\mathbf{W}_b^{\mathrm{dl}}\right)^{\mathsf{H}} \hat{\bm{v}}_b^{\mathrm{dl}}$.
We again ignore the distortion introduced by quantization for now and explain the details of designing the quantizer in Section \ref{sec:quantization}, i.e., assume ${\bm{v}}_b^{\mathrm{dl}}=\hat{\bm{v}}_b^{\mathrm{dl}}$,
 so that:
 \begin{equation}
    \bm{x}_b^{\mathrm{dl}} = \left(\mathbf{W}_b^{\mathrm{dl}}\right)^{\mathsf{H}} {\bm{v}}_b^{\mathrm{dl}} .
\label{eq:x_b_dl}
\end{equation}

Finally, the received signal at user $n$ can be written as a function of the transformation matrices by plugging \eqref{eq:v_dl} and \eqref{eq:x_b_dl} into  \eqref{eq:rn_dl} as:
\begin{equation}
    {y}_n^{\mathrm{dl}} 
    = \sum_{i=1}^N p_i^{\mathrm{dl}}\left(\bm{f}_n^{\mathrm{dl}}\right)^{\mathsf{H}}\bm{c}_i^{\mathrm{dl}}s_i^{\mathrm{dl}} + z_n^{\mathrm{dl}}
    .
\end{equation}
Thus, the achievable rate of user $n$ can be written as:
\begin{equation}
    R_{n}^{\mathrm{dl}}=\log \left( 1+\frac{p_n^{\mathrm{dl}}\left|\left(  \bm{c}_{n}^{\mathrm{dl}} \right) ^{\mathsf{H}}\bm{f}_{n}^{\mathrm{dl}} \right|^2}
    {\sum_{i\ne n}{p_i^{\mathrm{dl}}\left| \left( \bm{c}_{i}^{\mathrm{dl}} \right) ^{\mathsf{H}}\bm{f}_{n}^{\mathrm{dl}} \right|^2}+\sigma _{\mathrm{dl}}^{2}} \right) ,
    \label{eq:rate_R_dl_simple}
\end{equation}
where $\bm{f}_n^{\mathrm{dl}}$ is the $n$-th column of $\mathbf{{F}}^{\mathrm{dl}}$ representing the effective channel vector between user $n$ and all the RRHs.
It can be observed that the rate $R_n^{\mathrm{dl}}$ is a function of the transformation matrices $\left\{ \mathbf{W}_b^{\mathrm{dl}} \right\}_{b=1}^B$ through $\bm{c}_{n}^{\mathrm{dl}}$ and $\bm{f}_{n}^{\mathrm{dl}}$.
Thus, the transformation matrices $\left\{ \mathbf{W}_b^{\mathrm{dl}} \right\} _{b=1}^{B}$ can be optimized to maximize a downlink utility, e.g., the sum rate $\sum_{n}{R_n^{\mathrm{dl}}}$.

We remark that left-multiplying $\mathbf{W}^{\mathrm{dl}}$ by an invertible matrix $\mathbf{A}$, i.e., $\tilde{\mathbf{W}}^{\mathrm{dl}}=\mathbf{A} \mathbf{W}^{\mathrm{dl}}$ does not change the information theoretical maximum achievable rate of the downlink.
This is because the power constraint applies to the transmitted signal $\left\{ \bm{x}_b^{\mathrm{dl}} \right\}_{b=1}^B$, and the effect of the invertible matrix A can always be reversed without affecting the power constraints by replacing the original beamforming vector ${\bm{c}}_n^{\mathrm{dl}}$ with $\tilde{\bm{c}}_n^{\mathrm{dl}}=\left(\mathbf{A}^{\mathsf{H}}\right)^{-1}{\bm{c}}_n^{\mathrm{dl}}$.
Therefore, without loss of generality, we can impose the constraint that $\mathbf{W}_b^{\mathrm{dl}}$ is a semi-orthogonal matrix, i.e., $\mathbf{W}_b^{\mathrm{dl}}\left(\mathbf{W}_b^{\mathrm{dl}}\right)^\mathsf{H}=\mathbf{I}$, which can be satisfied using a QR decomposition step as done in the uplink. 
In effect, the semi-orthogonal transformation matrices at the RRHs define the suitable subspace in which the high-dimensional signal $\left\{ \bm{x}_b^{\mathrm{dl}} \right\}_{b=1}^B$ to be transmitted on the RRH antennas can be more compactly described.
The benefit of adding this constraint is that it can ensure the same signal power before and after the transformation, i.e., $\left\| \bm{x}_b^{\mathrm{dl}} \right\|^2 = \left\| {\bm{v}}_b^{\mathrm{dl}} \right\|^2$.
In this way, the power of the transmitted signals $\left\{\bm{x}_b^{\mathrm{dl}}\right\}_{b=1}^B$ at the RRHs will not be scaled by the transformation matrices and can be designed in the CP by adjusting the power factors $\left\{ p_n^{\mathrm{dl}}\right\}_{n=1}^N$.

\subsection{Unified Uplink and Downlink Problem Formulation}
It can be observed that the uplink and downlink system models are reciprocals of each other, and both the uplink and downlink achievable rates are functions of the transformation matrices.
Hence, assuming a quasi-static block-fading channel model, we can establish a unified problem formulation of designing uplink or downlink transformation matrices to maximize the utility within the channel coherence time in both the uplink and the downlink. 
As an example, this paper considers the system sum rate as the utility function. 
Note that with minor modifications, the framework can accommodate other utility functions as well.

With the system model described above, the uplink and downlink sum-rate maximization problem can be formulated in a unified way:
\begin{subequations}
    \begin{align}
        & \mathop \text{maximize} \limits_{\left\{ \mathcal{F} _b\left( \cdot \right) \right\} _{\forall b}} 
        & & R \triangleq {\mathbb{E}}\left[
        \sum_{n=1}^N {
        R_n\left(\left\{\mathbf{W}_b \right\}_{b=1}^B \right) } 
        \right] 
        \label{eq:prob_general_obj}\\
        &\text{subject to}
        & & \mathbf{W}_b=\mathcal{F} _b\left( \left\{ \mathbf{H}_b \right\} _{b=1}^{B} \right), \quad b=1,\dots,B , \\
        & & & \mathbf{W}_b \mathbf{W}_b^{\mathsf{H}}=\mathbf{I}, \quad b=1,\dots,B ,
        \label{eq:prob_general_gs} 
    \end{align}
\label{eq:prob_general_all}%
\end{subequations}
where the expectation in the objective function is over the distribution of channel matrices $\left\{ \mathbf{H}_b \right\} _{ b=1}^B$.
It is assumed that channel estimation has been done in advance so that perfect local CSI $\mathbf{H}_b$ is available at RRH $b$ for each coherence block.
Once the mapping functions $\left\{ \mathcal{F} _b\left( \cdot \right) \right\} _{\forall b}$ are optimized, they can be used to design the transformation matrices $\left\{\mathbf{W}_b \right\}_{b=1}^B$ from $\left\{\mathbf{H}_b \right\}_{b=1}^B$ at the beginning of each coherence block.

Solving the problem \eqref{eq:prob_general_all} is challenging in terms of both communication cost and computational complexity. 
Specifically, the optimal transformation matrices $\left\{\mathbf{W}_b \right\} _{b=1}^{B}$ should be designed jointly at the CP with the global CSI $\left\{\mathbf{H}_b \right\} _{b=1}^{B}$, but this requires significant communication overhead for transmitting the local CSI from the RRHs to the CP and transmitting the designed transformation matrices from the CP back to the RRHs through the fronthaul.
Moreover, given the global CSI, this problem is also computationally difficult to solve due to its nonconvexity. 

To reduce the communication overhead and the computational complexity, many existing works resort to suboptimal local CSI based algorithms.
For example, in \cite{liu20171,liu2015optimized,wiffen2019distributed} the problem of minimizing the reconstruction error of the received signal, i.e.,  $\mathbb{E}\left[ \left\| \bm{y}_b-\mathbf{W}_{b}^{\mathsf{H}}\mathbf{W}_b\bm{y}_b \right\| _{2}^{2} \right]$,
is proposed at each individual RRH for the uplink,
whose solution is given by the EVD method, also known as principal component analysis.
Intuitively, the EVD solution finds the strongest subspace of the local channel at each RRH, which is optimal in terms of minimizing the reconstruction MSE for the single RRH case.
However, for a C-RAN system with multiple RRHs, this EVD based approach is suboptimal since independent local metrics may not match the end-to-end system objective and finding the optimal solution needs the global CSI.

This paper proposes a novel data-driven approach to solve the problem \eqref{eq:prob_general_all}.
Instead of trying to solve \eqref{eq:prob_general_all} analytically, we generate many problem instances to train a set of DNNs to learn the optimized solutions empirically.
Further, to reduce the communication overhead while achieving good performance, this paper proposes an efficient distributed signaling architecture to coordinate the operations of the DNNs across the RRHs.
By carefully designing the neural network structure and the signaling scheme, near-optimal performance with significantly reduced communication overhead can be achieved.

\section{Two-Stage Design of Transformation Matrices} \label{sec:design_W}

The main contribution of this paper is a two-stage deep learning framework including an initialization stage and a refinement stage to solve the problem \eqref{eq:prob_general_all}.
Specifically, the transformation matrix at each RRH is first derived from the local CSI using a DNN in the first stage, and then refined iteratively using the gradient of the objective in the second stage as shown in Fig. \ref{fig:dnn_all}.
To reduce the communication overhead in the refinement stage, a low-dimensional signaling scheme is proposed for each iteration, and a meta-learning based GRU network is designed to reduce the number of iteration rounds.

\subsection{Stage One: Initialization Using Local CSI}
\label{sec:DNN_local_CSI}
In the first stage, each RRH uses a DNN to design its transformation matrix based on its local CSI.
Mathematically, RRH $b$ maps its local CSI $\mathbf{H}_b$ into the initial transformation matrix $\mathbf{W}_b^{(0)}$ according to:
\begin{equation}
    \mathbf{W}_{b}^{\left( 0 \right)}=\mathcal{P} _{\bm{\theta }_b}\left( \mathbf{H}_b \right)  ,
    \label{eq:mapping_local_CSI}
\end{equation}
where the superscript denotes the number of steps, and $\mathcal{P} _{\bm{\theta }_b}\left( \cdot \right) $ denotes the DNN with parameters $\bm{\theta}_b$.
We incorporate a normalization step as the final layer of the DNN where QR decomposition is adopted to satisfy the semi-orthogonal constraint $\mathbf{W}_b^{\mathrm{ul}} \left(\mathbf{W}_{b}^{\mathrm{ul}}\right)^{\mathsf{H}}=\mathbf{I}$.

The trained DNNs $\left\{\mathcal{P} _{\bm{\theta }_b}\left( \cdot \right) \right\}_{b=1}^B$ can already be deployed to design transformation matrices using local CSI.
Even without the second stage, this local CSI based DNN architecture can outperform the heuristic EVD method in \cite{liu20171} since the DNNs learn to optimize the end-to-end loss function instead of the intermediary MSE loss. 
Moreover, even though the global CSI is not accessible to each RRH, the local DNNs can implicitly learn to make use of the statistical distribution of the global CSI through training.

\subsection{Stage Two: Iterative Refinement Using Global Signaling}
\label{sec:stage2}

After the transformation matrices are initialized at the RRHs using the local CSI, a refinement process can be implemented to further improve the performance by exchanging some low-dimensional signals between the RRHs and the CP iteratively.
To enhance the efficiency of the iterative algorithm, we propose a low-dimensional signaling strategy for each iteration and a GRU based meta-learning framework to accelerate the convergence speed.

\begin{table}[t]
\small
\caption{Communication overhead between RRH $b$ and the CP.}

\centering
\begin{tabular}{|l|c|c|c|}
\hline
Methods 
& \begin{tabular}[c]{@{}l@{}}Uplink\\ Transmission\end{tabular} 
& \begin{tabular}[c]{@{}l@{}}Downlink\\ Transmission\end{tabular} 
& \begin{tabular}[c]{@{}l@{}}Total \\ Communication \\Overhead \end{tabular}  \\ 

\hline

\begin{tabular}[c]{@{}l@{}}Local CSI\\ (EVD/DNN)\end{tabular} 
& $\mathbf{F}_b$
& --
&  $KN$ \\ 

\hline

\begin{tabular}[c]{@{}l@{}}Two-stage\\ DNN\end{tabular} 
& 
\begin{tabular}[c]{@{}c@{}}
$ \mathbf{F}_{b}^{\left( t \right)}$
\\ $t=0, \dots, T$
\end{tabular}
& 
\begin{tabular}[c]{@{}c@{}}
{$ \left. \frac{\partial R}{\partial \mathbf{F}_b} \right|_{\mathbf{F}_b=\mathbf{F}_{b}^{\left( t \right)}}$}
\\ {$t=0, \dots,$} \\{$~~~~~~~ T-1$}
\end{tabular}

& $(2T+1)KN$ \\ 

\hline

Global CSI 
& $\mathbf{H}_b$
& $\mathbf{W}_b$
&  $MN+KM$\\ 

\hline
\end{tabular}
\label{table}
% \vspace{-0.5cm}
\end{table}

\subsubsection{Low-Dimensional Signaling Scheme}

A straightforward signaling scheme would involve the following steps: RRH $b$ sends the transformation matrix $ \mathbf{W}_{b}$ to the CP;
the CP calculates the gradient
$\partial R/\partial \mathbf{W}_b$ and sends it back to RRH $b$ to update the transformation matrix $ \mathbf{W}_{b}$ using the gradient descent (GD) algorithm.
However, this naive scheme requires global CSI matrices $\left\{ \mathbf{H}_b \right\} _{b=1}^{B}$ at the CP to compute the gradient ${\partial R}/{\partial \mathbf{W}_{b}}$.

\begin{figure*}[t]
    \centering
    \subfloat[The uplink and downlink low-dimensional signaling scheme.]{%
      \includegraphics[width=0.62\textwidth]{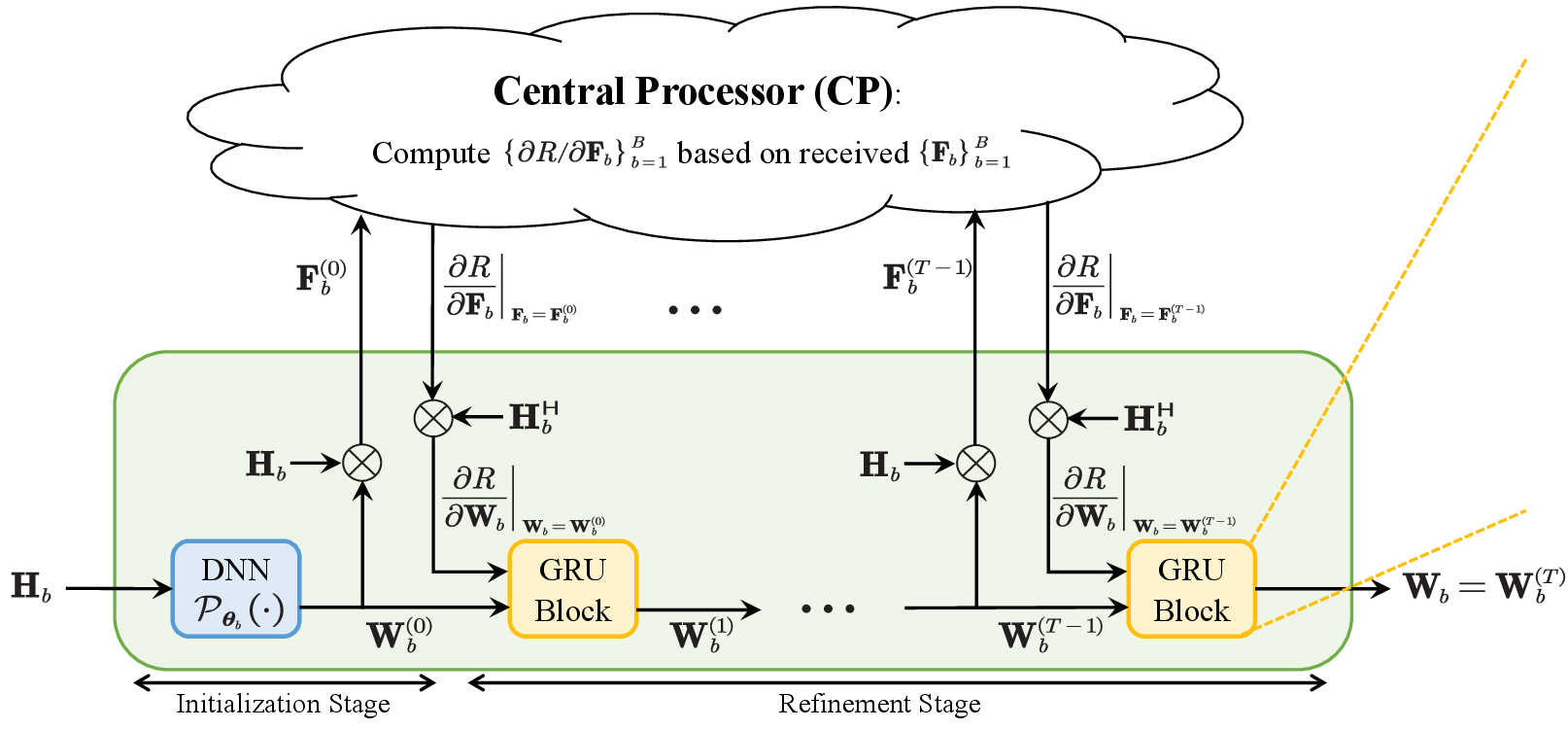}%
      \label{fig:dnn_module}%
    }
    \subfloat[The GRU block for updating the transformation matrix at RRH $b$ in the $t$-th iteration.]{%
      \raisebox{15pt}{\includegraphics[width=0.35\textwidth]{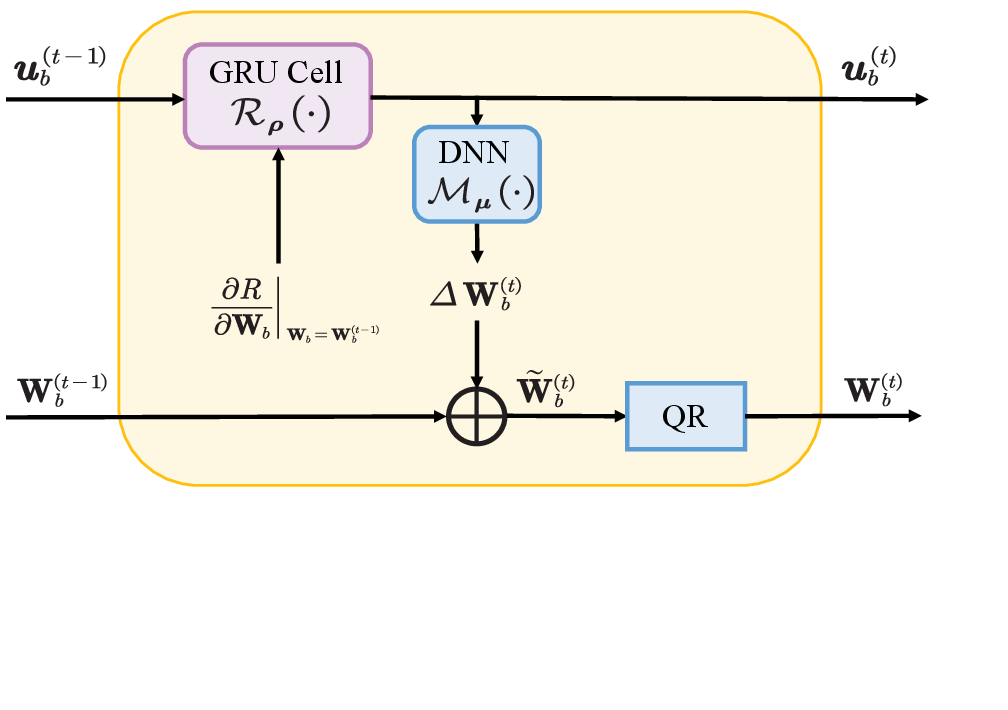}}%
      \label{fig:gru}%
    }
    \caption{A block diagram of the proposed two-stage meta-learning algorithm for designing transformation matrices.}
    \label{fig:dnn_all}
\end{figure*}

In this paper, we observe that the system sum rate $R$ is only a function of effective CSI matrices;
thus, we have:
\begin{equation}
    \frac{\partial {R}}{\partial \mathbf{W}_{b}}=\frac{\partial {R}}{\partial \mathbf{F}_{b}}\mathbf{H}_{b}^{\mathsf{H}} ,
\label{eq:grad_W_from_grad_F}
\end{equation}
which implies that the gradient with respect to $\mathbf{W}_b$ can be recovered at RRH $b$ based on the gradient with respect to $\mathbf{F}_b$ and the local CSI $\mathbf{H}_b$.
Therefore, we propose a signaling scheme as shown in Fig. \ref{fig:dnn_module}. 
Specifically, in the $t$-th iteration, RRH $b$ sends the CP its current low-dimensional effective CSI $\mathbf{F}_b^{(t-1)}$.
After collecting the effective CSI matrices from all the RRHs, the CP computes the gradient $\partial R/\partial \mathbf{F}_b$ evaluated at ${\mathbf{F}_b=\mathbf{F}_{b}^{\left( t-1 \right)}}$ and transmits it back to RRH $b$.
Then, RRH $b$ recovers the gradient $\partial R/\partial \mathbf{W}_{b}$ evaluated at ${\mathbf{W}_b=\mathbf{W}_{b}^{\left( t-1 \right)}}$ according to \eqref{eq:grad_W_from_grad_F} and performs the GD update given by:
\begin{equation}
    \tilde{\mathbf{W}}_{b}^{\left( t \right)}=\mathbf{W}_{b}^{\left( t-1 \right)}+\alpha _t \frac{\partial R}{\partial \mathbf{W}_{b}}\bigg|_{{\mathbf{W}_b=\mathbf{W}_{b}^{\left( t-1 \right)}}} ,
    \label{eq:gradient_descent_W}
\end{equation}
where $\alpha _t$ is the step size.
At the end of each iteration,
we take the QR decomposition $\left( \tilde{\mathbf{W}}_b^{(t)}\right) ^{\mathsf{H}}=\mathbf{Q} \mathbf{R}$ and set the $\mathbf{Q}^\mathsf{H}$ matrix as the orthogonalized $\mathbf{W}_b^{(t)}$.
The final transformation matrices are produced after $T$ iterations, i.e., we set $\mathbf{W}_b = \mathbf{W}_{b}^{\left( T \right)}$.

This paper quantifies the amount of communication overhead by the number of entries in the signaling matrices. 
For each iteration of the proposed scheme, both the uplink and downlink signaling are $K\times N$ dimensional, consuming a total of $ (2T+1)KN$ overhead per RRH for $T$ refinement iterations and one final uplink transmission.
Comparatively, the global CSI based approach requires the transmission of the $M \times N$ dimensional full CSI in the uplink, and the $K \times M$ dimensional transformation matrix in the downlink for each RRH.
Since $K$ is much smaller than the number of antennas $M$, the amount of overhead can be significantly reduced by the proposed low-dimensional iterative signaling scheme provided that the number of iterations is small. 
The communication overhead of different methods is summarized in Table \ref{table}.

\subsubsection{Reducing Signaling Rounds via Meta-Learning}

To reduce the number of communication rounds $T$, it is crucial to design an algorithm that can converge quickly.
However, GD in general has a slow convergence rate since it only uses the gradient information in each time step.
Moreover, the optimal step size for each iteration is difficult to choose.
To accelerate the convergence speed, this paper proposes a meta-learning based GRU network that learns the update step based on the current and historical gradient information \cite{andrychowicz2016learning, cho2014properties, xia2022metalearning}. 
Specifically, as shown in Fig. \ref{fig:gru}, in the $t$-th refinement iteration at RRH $b$, the GRU cell takes the previous hidden state vector $\bm{u}_b^{(t-1)}$ and the gradient $\partial R/\partial \mathbf{W}_{b}$ evaluated at the previous transformation matrix $ \mathbf{W}_b^{(t-1)}$ as inputs, and outputs the new hidden state vector $\bm{u}_b^{(t)}$ according to:
\begin{equation}
    \bm{u}_{b}^{\left( t \right)}=\mathcal{R} _{\bm{\rho }}\left( \bm{u}_{b}^{\left( t-1 \right)}, \left.  \frac{\partial R}{\partial \mathbf{W}_b} \right|_{\mathbf{W}_b=\mathbf{W}_{b}^{\left( t-1 \right)}} \right)   ,
    \label{eq:compute_cbt}
\end{equation}
where $\mathcal{R} _{\bm{\rho }}\left( \cdot \right) $ denotes the GRU hidden state update function with parameters $\bm{\rho }$.
Using another DNN $\mathcal{M} _{\bm{\mu }}\left(\cdot \right)$ with parameters $\bm{\mu }$, the update term $\Delta \mathbf{W}_b^{(t)}$ is mapped from the hidden state vector $\bm{u}_b^{(t)}$ according to:
\begin{equation}
    \Delta \mathbf{W}_{b}^{\left( t \right)}=\mathcal{M} _{\bm{\mu }}\left( \bm{u}_{b}^{\left( t \right)} \right)  .
    \label{eq:delta_W}
\end{equation}
Finally, the transformation matrix $ \mathbf{W}_{b}^{(t-1)}$ is updated as:
\begin{equation}
    \mathbf{\tilde{W}}_{b}^{\left( t \right)}=\mathbf{W}_{b}^{(t-1)}+\Delta \mathbf{W}_{b}^{(t)} ,
    \label{eq:update}
\end{equation}
followed by the QR decomposition step.
The overall algorithm is summarized in \textbf{Algorithm \ref{alg}}.

We remark that the proposed meta-learning based method offers several advantages over the traditional GD approach. 
Firstly, the GRU can leverage historical gradient information stored in the hidden state vector to inform the current update step of the transformation matrix, which effectively introduces momentum to the optimization task.
Secondly, the GRU learns an efficient update rule for this specific problem from the training data rather than relying on manually designed general update rules.
Lastly, while the step size is hard to configure for GD, it is already incorporated in the update step $\Delta \mathbf{W}_{b}^{\left( t \right)}$ in the GRU meta-learning block and is not required to be configured manually.
Overall, these advantages enable the proposed meta-learning based approach to converge much more quickly as compared to traditional GD, as will be illustrated subsequently.

\subsection{Neural Network Training}

Under the proposed framework, each RRH $b$ is equipped with an initial DNN with parameters $\bm{\theta}_b$, serving as the component of stage one, and a GRU block with parameters $\bm{\rho}$ and $\bm{\mu}$, serving as the component of stage two.
Since the GRU block acts as a meta-learner, i.e., an optimizer to update the transformation matrices, the same GRU block is shared for all the RRHs and is reused for $T$ times as shown in Fig. \ref{fig:dnn_module}.
The overall two-stage network is trained in an end-to-end manner to optimize the network parameters $\left\{ \left\{ \bm{\theta }_b \right\} _{b=1}^{B},\bm{\rho},\bm{\mu} \right\} $.

We notice that the true objective function to be minimized, i.e., the final negative sum rate,
is an immediate function of only the transformation matrices in the final update step.
This makes the gradient backpropagation for the earlier steps inefficient in the neural network training stage.
Therefore, we instead minimize the accumulated sum rate $- \sum_{t=0}^T{R\left( \left\{\mathbf{W}_{b}^{\left( T \right)} \right\}_{b=1}^B \right)}$ through update steps in the training phase as suggested in \cite{andrychowicz2016learning}.

\begin{figure}[t]
\begin{algorithm}[H]
\setstretch{0.8}
  \caption{Proposed two-stage meta-learning algorithm to design transformation matrices}
  \begin{algorithmic}[1]
    \State \textit{\# Stage 1: Initialization using local CSI}
        \State Initialize $\mathbf{W}_{b}^{\left( 0 \right)} $ via \eqref{eq:mapping_local_CSI}, $b=1, \dots B$
    \State \textit{\# Stage 2: Iterative refinement using global signaling}
      \State Initialize GRU hidden state vector $\bm{s}_{b}^{\left( 0 \right)}$
      \For{$t = 1:T$ }    
        \State RRH $b$ sends effective CSI $ \mathbf{F}_{b}^{\left( t-1 \right)}$ to the CP,  $b=1, \dots B$
        \State CP computes gradient $ \partial R/\partial \mathbf{F}_b|_{\mathbf{F}_b=\mathbf{F}_{b}^{\left( t-1 \right)}}$ 
        \Statex \qquad and sends it back to RRH $b$, $b=1, \dots B$
        \For{each RRH~$b=1, \dots B$}
            \State Compute gradient $\partial R/\partial \mathbf{W}_b|_{\mathbf{W}_b=\mathbf{W}_{b}^{\left( t-1 \right)}}$ via \eqref{eq:grad_W_from_grad_F}
            \State Update hidden state vector $\bm{s}_{b}^{\left( t \right)}$ via \eqref{eq:compute_cbt}
            \State Compute update term $\Delta \mathbf{W}_{b}^{\left( t \right)}$ via \eqref{eq:delta_W}
            \State Update $\mathbf{W}_{b}^{\left( t \right)}$ via \eqref{eq:update} and QR decomposition
        \EndFor 
    \EndFor    
    
    \State Set $\mathbf{W}_b = \mathbf{W}_b^{(T)}$
    
  \end{algorithmic}

  \label{alg}

\end{algorithm}
\end{figure}

\section{Uniform Quantization Scheme} \label{sec:quantization}

In this section, we design a $Q_b$-bit uniform scalar quantizer, denoted as $\mathcal{Q}_b\left(\cdot \right)$, for compressing the low-dimensional signal $\bm{v}_b$, where the value of $Q_b$ is determined by the fronthaul capacity of the link connecting RRH $b$ and the CP. 
This quantizer is deployed both at RRH $b$ and the CP.
Specifically, RRH $b$ compresses the signal $\bm{v}_b^\mathrm{ul}$ in the uplink, while the CP compresses the signal $\bm{v}_b^\mathrm{dl}$ in the downlink.
To allow for a tractable analysis of the achievable rate, we model the quantized signal as:
\begin{equation}
    \hat{\bm{{v}}}_{b} = \bm{v}_{b} + \bm{e}_{b} ,
\end{equation}
where $\bm{e}_{b}$ is the quantization error with a diagonal covariance matrix $\mathbf{D}_b \in \mathbb{C}^{K \times K}$ whose $k$-th diagonal entry $D_{bk}$ is given by the quantization noise variance at the $k$-th dimension.
Note that $e_{bk}$ values are independent across RRHs due to the independent quantization among the RRHs.

The uniform quantization of the $k$-th dimension of the complex vector $\bm{v}_b$, i.e.,  ${v}_{bk}={v}_{bk}^{\mathrm{I}}+j\cdot {v}_{bk}^{\mathrm{Q}}$ can be implemented via separate in-phase/quadrature (I/Q) quantization of the I-branch real symbol ${v}_{bk}^{\mathrm{I}}$ and the Q-branch real symbol ${v}_{bk}^{\mathrm{Q}}$. 
A key parameter in the design of the uniform quantization scheme is to choose the dynamic range.
This paper proposes to use a scaled version of the expected standard deviation $\sigma _{bk}$ of the symbol $v_{bk}$.
That is, the uniform quantization dynamic range of both ${v}_{bk}^{\mathrm{I}}$ and ${v}_{bk}^{\mathrm{Q}}$ is set as:
\begin{equation}
    d_{bk}=\frac{\gamma}{\sqrt{2}} \sigma _{bk} ,
    \label{eq:quant_clipping}
\end{equation}
where $\gamma$ is a constant scaling factor.
Accordingly, we apply a clipping function to enforce the quantized symbol into the dynamic range $\pm d_{bk}$ and set the quantization step size as:
\begin{equation}
    \Delta _{bk}=\frac{2d_{bk}}{2^{Q_{bk}}} .
\end{equation}
Further, the quantized value of ${v}_{bk}$ is given by:
\begin{equation}
    \hat{v}_{bk}=\Delta _{bk}\lceil \frac{v_{bk}}{\Delta _{bk}} \rceil -\frac{\Delta _{bk}}{2} .
\end{equation}
The I/Q quantization error ${e}_{bk}^{\mathrm{I}}$ and ${e}_{bk}^{\mathrm{Q}}$ can be modeled with uniform distribution as: 
\begin{equation}
    e_{bk}^{\mathcal{X}}\sim \mathcal{U} \left( -\frac{d_{bk}}{2^{Q_{bk}}},\frac{d_{bk}}{2^{Q_{bk}}} \right),  \quad \mathcal{X} \in \left\{ \mathrm{I},\mathrm{Q}\right\} ,
    \label{eq:uniform_quant_noise}
\end{equation}
where $Q_{bk}=Q_b / K$ is the number of bits assigned to ${v}_{bk}^{\mathrm{I}}$ and ${v}_{bk}^{\mathrm{Q}}$.
Therefore, the MSE distortion between the continuous-valued symbol ${v}_{bk}$ and the quantized symbol $\hat{v}_{bk}$ can be written as:
\begin{equation}
    D_{bk}=\frac{\gamma^2 \sigma_{bk}^{2}}{3\cdot 2^{2Q_{bk}}} .
    \label{eq:D_bk}
\end{equation}

Since $D_{bk}$ in \eqref{eq:D_bk} is differentiable with respect to $\sigma _{bk}^{2}$ and further over $\mathbf{W}_b$, we can potentially incorporate the quantization step directly into the neural network training process, i.e., 
assuming that the quantization error is AWGN with noise power given by \eqref{eq:D_bk} and including the quantization noise terms in the sum rate objective with a trainable scaling factor $\gamma$. 
However, through simulations, we observe that considering quantization during training can only bring marginal benefits as compared to the scheme ignoring quantization during training and adding quantization afterwards.
In other words, the transformation matrices designed using the proposed scheme in Section \ref{sec:design_W} in which the quantization effects are ignored already lead to excellent performance. 
Moreover, optimizing the rate incorporating quantization requires us to train an individual DNN for each fronthaul rate constraint, which significantly increases training complexity.
Thus, we do not incorporate quantization effects into the training process.

Finally, we remark on the effect of selecting the clipping range scaling factor $\gamma$.
If $\gamma$ is too large, it will result in coarser quantization and thus higher quantization noise. 
Conversely, if $\gamma$ is too small, many entries in vector $\bm{v}_{bk}$ will exceed the quantization clipping range, leading to higher distortion. 
To deal with the overflow effect in the objective function, we remove the overflow dimensions from the received signal and use other signal entries in the cooperative cluster to decode the user symbols in the CP. 
In practice, $\gamma$ can be found with a simple line search.
Simulation results show that the system sum rate objective is not sensitive to the choice of $\gamma$ as long as it is within a reasonable range.

\section{Uplink sum-rate maximization Problem with User-Centric Clustering}\label{sec:ul}

As an illustrative example of the proposed algorithm, we evaluate the performance of using the proposed framework in Sections \ref{sec:design_W} and \ref{sec:quantization} to design the fronthaul compression strategy for the uplink channel model  \eqref{eq:y_b_ul}.
While it is previously assumed in Section \ref{sec:model} that all the RRHs can jointly serve all the users,  this is not practical for large networks due to the high computational complexity at the CP.
In this section, we adopt a user-centric clustering scheme \cite{zhu2018stochastic}, where each user is 
served by a cluster of RRHs of limited size, and clusters for different users may overlap. 
Taking this clustering scheme into account, we rewrite the uplink system model and present the simulation results.

\subsection{System Model}

In the C-RAN uplink, the received signal $\bm{y}_b^{\mathrm{ul}} \in \mathbb{C}^M$ at RRH $b$ is given by \eqref{eq:y_b_ul} and is then transformed into the low-dimensional signal $\bm{v}_b^{\mathrm{ul}} \in \mathbb{C}^K$ according to \eqref{eq:v_b_ul} with the semi-orthogonal transformation matrix $\mathbf{W}_b^{\mathrm{ul}}$ designed using the two-stage algorithm in Section \ref{sec:design_W}.
The effective channel between RRH $b$ and all the users is defined as in \eqref{eq:F_b_ul}.
Next, the dimension-reduced signal $\bm{v}_b^{\mathrm{ul}}$ is quantized into the digital signal $\hat{\bm{v}}_b^{\mathrm{ul}}$ using the $Q_b$-bit uniform scalar quantizer $\mathcal{Q}_b^{\mathrm{ul}}\left(\cdot \right)$ introduced in Section \ref{sec:quantization} and then forwarded onto the digital fronthaul channel.

At the CP, the user signals are recovered based on the compressed signal vectors from the RRHs within its serving cluster.
Specifically, each user is associated with its strongest RRHs. 
We use $\Theta _n$ to denote the serving cluster of RRHs for user $n$ and use $\left| \Theta _n \right|$ to denote the cluster size. 
Then, the  received signal of the RRH cluster $\Theta_n$, represented as $\bm{\bar{v}}_n^{\mathrm{ul}}=\left[ \cdots \left(\bm{v}_{b}^{\mathrm{ul}} \right)^{\mathsf{H}}\cdots \right] _{b\in \Theta _n}^{\mathsf{H}}$, can be written as:
\begin{equation}
    \bm{\bar{v}}_n^{\mathrm{ul}}=\mathbf{\bar{F}}_n^{\mathrm{ul}}\bm{x}^{\mathrm{ul}}+\mathbf{\bar{W}}_n^{\mathrm{ul}}\bm{\bar{z}}_n^{\mathrm{ul}} + \bar{\bm{e}}_n^{\mathrm{ul}} ,
    \label{eq:v_n_bar}
\end{equation}
where $\mathbf{\bar{F}}_n^{\mathrm{ul}}=\left[ \cdots \left( \mathbf{F}_{b}^{\mathrm{ul}}\right)^{\mathsf{H}}\cdots \right] _{b\in \Theta _n}^{\mathsf{H}}\in \mathbb{C} ^{\left| \Theta _n \right|K\times N}$ is the  effective channel matrix after compression, 
$\mathbf{\bar{W}}_n^{\mathrm{ul}}\in \mathbb{C} ^{\left| \Theta _n \right|K\times \left| \Theta _n \right|M}$ is the block-diagonal transformation matrix for the RRH cluster $\Theta_n$,
$\bm{\bar{z}}_n^{\mathrm{ul}}=\left[ \cdots \left(\bm{z}_{b}^{\mathrm{ul}}\right)^{\mathsf{H}}\cdots \right] _{b\in \Theta _n}^{\mathsf{H}}\in \mathbb{C} ^{\left| \Theta _n \right|M}$ is the  noise vector,
and $\bm{\bar{e}}_n^{\mathrm{ul}}=\left[ \cdots \left(\bm{e}_{b}^{\mathrm{ul}} \right)^{\mathsf{H}}\cdots \right] _{b\in \Theta _n}^{\mathsf{H}}\in \mathbb{C} ^{\left| \Theta _n \right|K}$ is the  quantization error vector.
We use $\bm{f}_{n,i}^{\mathrm{ul}}=\mathbf{\bar{W}}_n^{\mathrm{ul}}\left[ \cdots \left(\bm{h}_{bi}^{\mathrm{ul}}\right)^{\mathsf{H}}\cdots \right] _{b\in \Theta _n}^{\mathsf{H}} \in \mathbb{C} ^{\left| \Theta _n \right|K}$ to represent the effective channel vector between user $i$ and the RRHs in $\Theta_n$.
Finally, the CP designs the 
linear MMSE
receive beamformer ${\bm{c}}_n^{\mathrm{ul}}$ using the effective CSI between the RRHs within the cluster $\Theta_n$ and all the users (while ignoring the effect of quantization) as in \eqref{eq:cn_ul_W}, where $\mathbf{{F}}^{\mathrm{ul}} $ and $\bm{f}_{n}^{\mathrm{ul}}$ are replaced by $\mathbf{\bar{F}}_n^{\mathrm{ul}}$ and $\bm{f}_{n,n}^{\mathrm{ul}}$, respectively.
Applying the beamformers $\left\{ \bm{c}_n^{\mathrm{ul}} \right\}_{n=1}^N$ to the received signal $\left\{ \bar{\bm{v}}_n^{\mathrm{ul}} \right\}_{n=1}^N$ and assuming that the quantization noise is AWGN, we obtain the achievable rate of user $n$ as:
\begin{equation}
\begin{aligned}
    &\tilde{R}_{n}^{\mathrm{ul}}= \\
    &\log \left( 1 
    +\frac{p_{n}^{\mathrm{ul}} | ( \bm{c}_{n}^{\mathrm{ul}} ) ^{\mathsf{H}}\bm{f}_{n,n}^{\mathrm{ul}} |^2}{\sum\limits_{\substack {i\ne n}}{p_{i}^{\mathrm{ul}}| ( \bm{c}_{n}^{\mathrm{ul}} ) ^{\mathsf{H}}\bm{f}_{n,i}^{\mathrm{ul}} |^2+\sigma _{\mathrm{ul}}^{2}\left\| \bm{c}_{n}^{\mathrm{ul}} \right\| ^2}+| ( \bm{c}_{n}^{\mathrm{ul}} ) ^{\mathsf{H}}\mathbf{\bar{D}}_{n}^{\mathrm{ul}}\bm{c}_{n}^{\mathrm{ul}} |^2} \right)  ,
\end{aligned}
\label{eq:rate_R_ori_ul}
\end{equation}
where 
$\bar{\mathbf{D}}_n^{\mathrm{ul}} \in \mathbb{C} ^{\left| \Theta _n \right|K\times \left| \Theta _n \right|K}$ is the block-diagonal covariance matrix of the quantization error 
vector $\bar{\bm{e}}_{n}^{\mathrm{ul}}$ with block-diagonal entries being $\mathbf{D}_b^{\mathrm{ul}}$, for $b\in \Theta_n$.
Note that the rate $R_n^{\mathrm{ul}}$ with the quantization error ignored, which is used as the objective during the training stage of the neural network, can be expressed by excluding $\left| \left( \bm{c}_{n}^{\mathrm{ul}} \right) ^{\mathsf{H}}\mathbf{\bar{D}}_{n}^{\mathrm{ul}}\bm{c}_{n}^{\mathrm{ul}} \right|^2$ from \eqref{eq:rate_R_ori_ul}.

\subsection{Neural Network Implementation Details}\label{sec:dnn_details_ul}

Although it is possible for each RRH to learn its own DNN as the mapping function from its local CSI to the initial transformation matrix, we train a common DNN shared across all the RRHs to save the training memory and complexity because the multi-cell wrap-around simulation topology adopted in this paper is symmetric.
The shared initial DNN has three fully-connected layers with hidden layers of width $\left[2048, 512 \right]$ and has $\tanh$ nonlinearity.
In the second stage of the proposed framework, the GRU and DNN in each iteration block are reused for every iteration and every RRH to save memory as mentioned above.
The size of the GRU hidden unit is $2KM$, and the DNN has one hidden layer of size $2KM$. 
We implement the deep learning models in PyTorch \cite{paszke2019pytorch} and train them using the Adam optimizer \cite{kingma2014adam} with an initial learning rate of  $10^{-4}$ and the \textit{ReduceLROnPlateau} scheduler.

\subsection{Numerical Results}\label{sec:sim_ul}
\subsubsection{Simulation Setup}
We evaluate the performances of the proposed scheme in the uplink sum-rate maximization problem with the 19-cell wrap-around cellular network simulation topology. 
The simulation considers a dense urban network, where the distance between two neighbouring RRHs is $150$m and the height of an RRH is $30$m. 
We perform simulations under two scenarios: the number of antennas $M=8$ and $M=32$, respectively.
We set $\left| \Theta _n \right|=7$ for $M=8$, and $\left| \Theta _n \right|=17$ for $M=32$.
In both cases, the compression dimension $K$ is set to $2$.
During each scheduling time slot, $2$ users are uniformly generated in each cell, and the transmit power of each user is $23$dBm.
The system bandwidth is $20$MHz and the background noise level is $-169$dBm/Hz. 
A signal-to-interference-plus-noise ratio (SINR) gap of $6$dB is considered to account for the coding and modulation scheme used in practice.
We assume that the carrier frequency is $2.9$GHz and the channel follows Rayleigh fading with the path loss $\beta = 41.74+29\log 10\left( d \right) $ \cite{sun2016investigation}.
Simulations are performed in the real field for simplicity.

The performance of the proposed scheme is compared against the following benchmarks:
\begin{itemize}
    \item \textbf{Single-cell processing.} This corresponds to the setting of a traditional non-cooperative cellular MIMO network, where each base station employs the linear MMSE beamformer \cite{joham2005linear} for the users in its own cell using the local CSI.
    
    \item \textbf{EVD using local CSI \cite{liu20171}.} The rows of the transformation matrix $\mathbf{W}_b$ for RRH $b$ are derived by taking the eigenvectors corresponding to the $K$ largest eigenvalues of $\mathbf{H}_b\mathbf{H}_{b}^{\mathsf{H}}$. The effective channel $\mathbf{F}_b = \mathbf{W}_b \mathbf{H}_b$ is forwarded to the CP to design receive beamformers. 

    \item \textbf{Modified EVD using local CSI and large scale fading statistics \cite{wiffen2021distributed}.}
    Assuming that the RRHs have access to the statistics of global CSI, this modified EVD approach designs the transformation matrix $\mathbf{W}_b^{\mathrm{ul}}$ by taking the first $K$ eigenvectors of $\mathbf{H}_b^{\mathrm{ul}}\left( p_n^{\mathrm{ul}}\sum_{j\ne b}{\mathbf{\Psi }_j+\sigma _{\mathrm{ul}}^{2}\mathbf{I}} \right) ^{-1}\left(\mathbf{H}_{b}^{\mathrm{ul}}\right)^{\mathsf{H}}$, where $\Psi_j$ is a diagonal matrix whose $n$-th diagonal entry is given by $\mathbb{E} \left[ \left\| \bm{h}_{jn}^{\mathrm{ul}} \right\| ^2 \right] =\beta _{jn}M$.
    
    \item \textbf{GD using global CSI.} 
    To reap the full cooperation gain, each RRH forwards its local CSI to the CP, where the transformation matrices for all the RRHs are jointly optimized using the global CSI with GD. 
    The designed matrices are then transmitted back to each RRH.
    
    \item \textbf{DNN using local CSI (first stage) + GD.} 
    The second stage of the proposed network is replaced with generic GD, which has the same signaling scheme as in  Section \ref{sec:stage2} but uses the update rule in \eqref{eq:gradient_descent_W}.
    The constant step size is tuned manually so that GD can achieve the best performance within the number of iterations allowed.
\end{itemize}

\begin{figure}[t]
    \centering
    {\includegraphics[width=8cm]{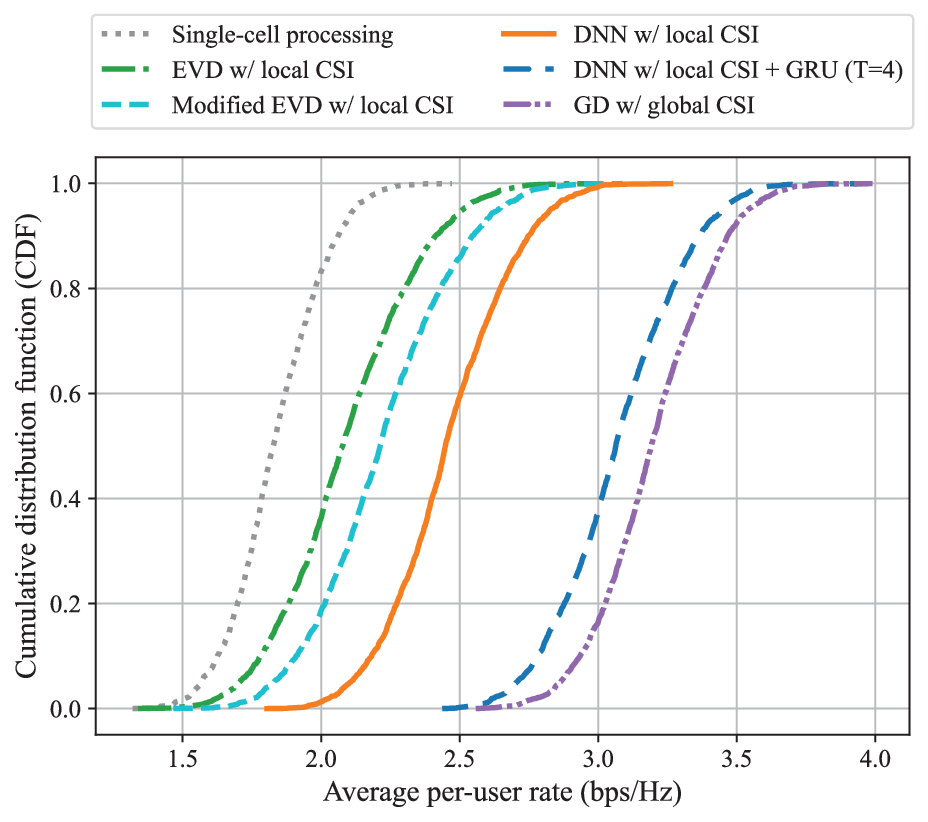}}
    \caption{Empirical CDF curve of uplink average per-user rate ($M=8$).}
    \label{fig:cdf_M8_ul}
    % \vspace{-0.3cm}
\end{figure}

\begin{figure}[t]
    \centering
    {\includegraphics[width=8cm]{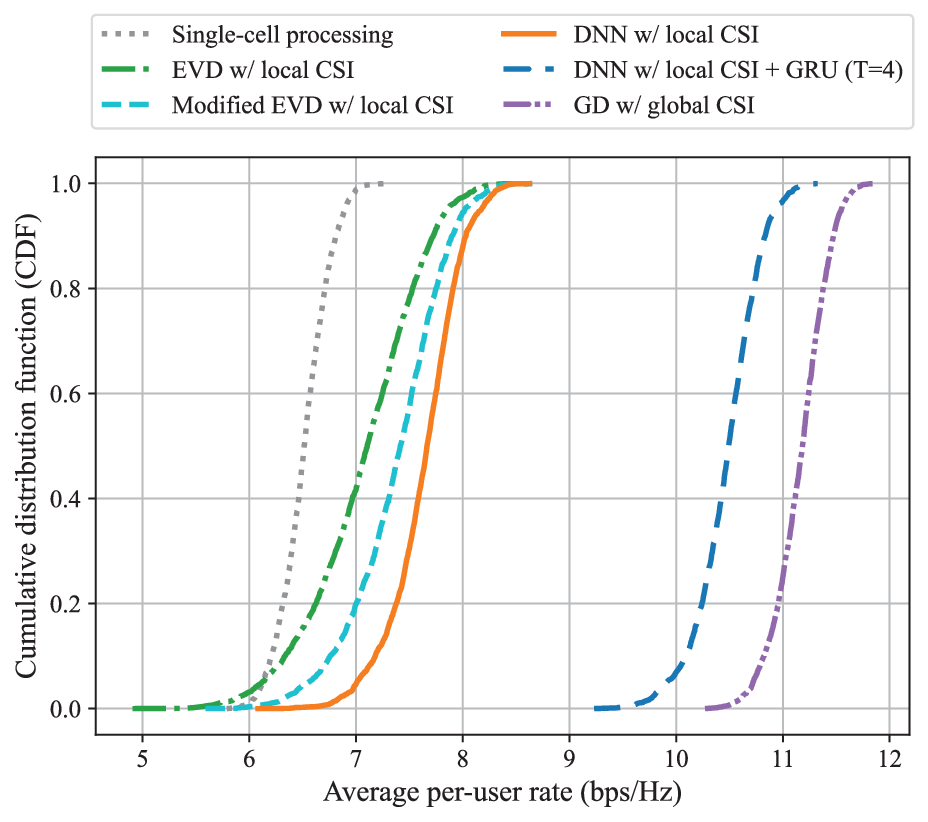}}
    \caption{Empirical CDF curve of uplink average per-user rate ($M=32$).}
    \label{fig:cdf_M32_ul}
    % \vspace{-0.5cm}
\end{figure}

\begin{figure}[t]
    \centering
    {\includegraphics[width=8cm]{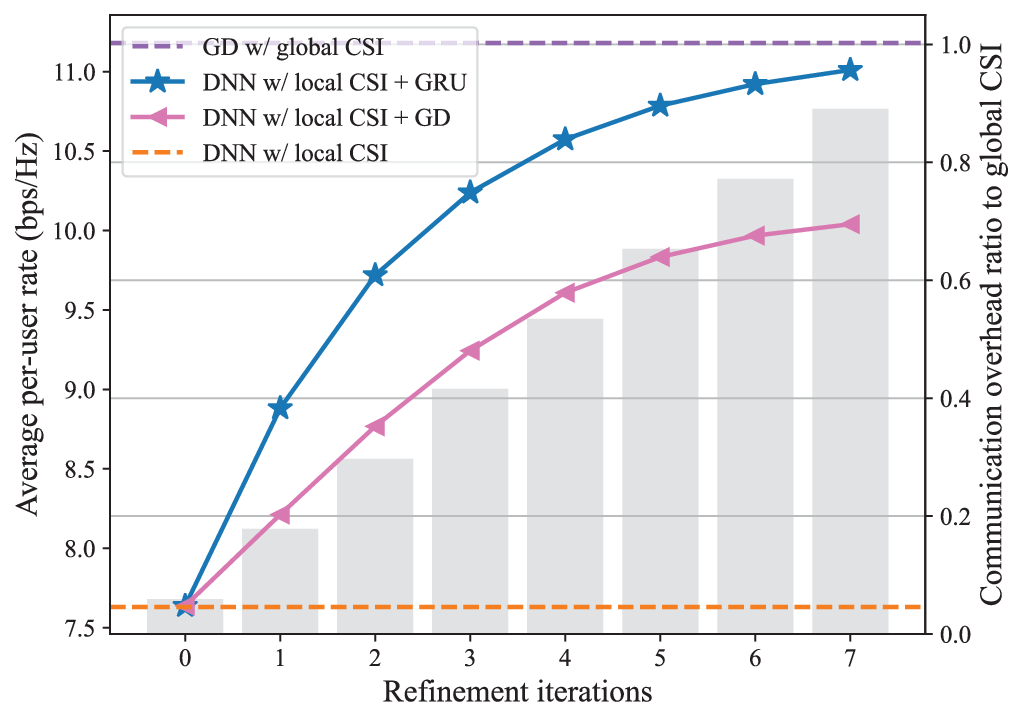}}
    % \vspace{-0.3cm}
    \caption{Left vertical axis: convergence curve for the uplink per-user rate. Right vertical axis: bar graph indicating the communication overhead ratio of the proposed iterative scheme to the global CSI based method ($M=32$).}
    % \vspace{-0.5cm}
    \label{fig:convergence_ul}
\end{figure}

\begin{figure}[t]
    \centering
    {\includegraphics[width=8cm]{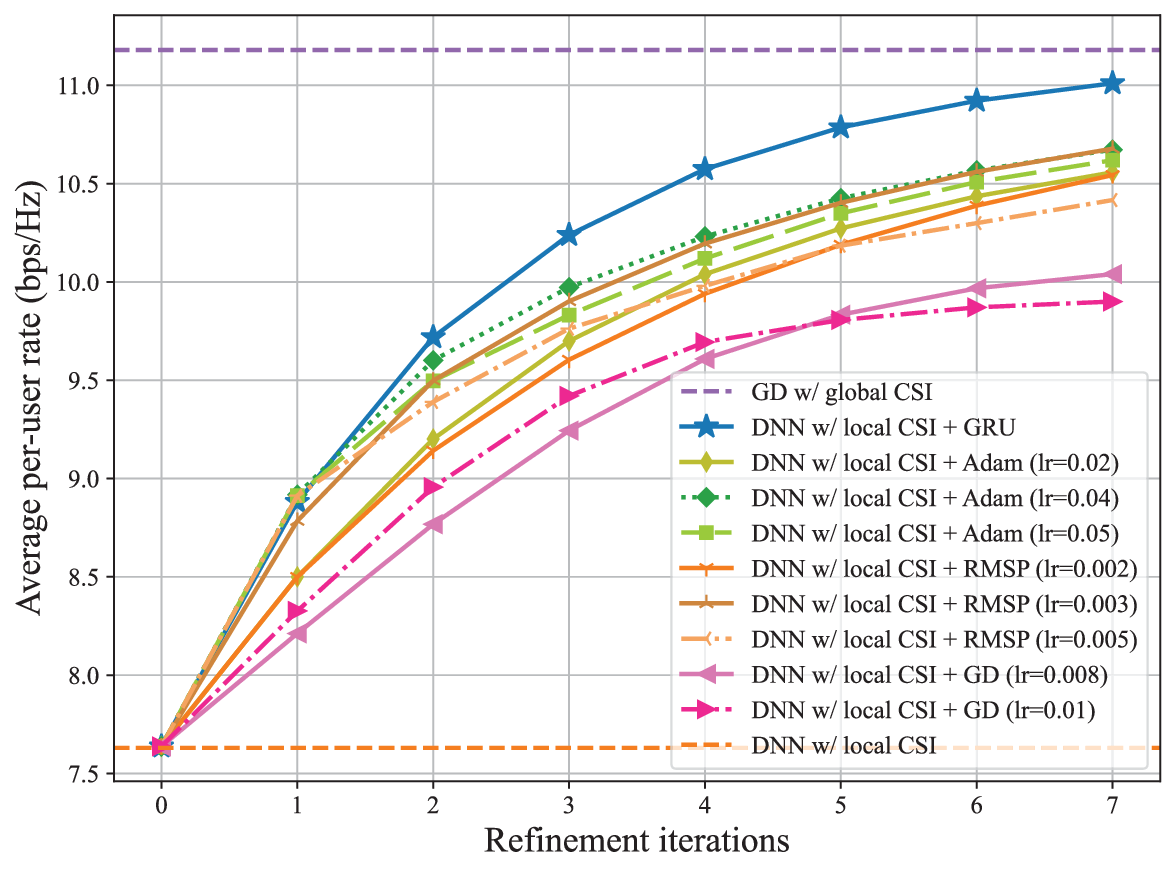}}
    % \vspace{-0.3cm}
    \caption{Convergence curve for the uplink per-user rate using different optimizers and step sizes ($M=32$).}
    % \vspace{-0.5cm}
    \label{fig:convergence_ul_adam}
\end{figure}

\subsubsection{Performance of the Proposed Two-Stage Algorithm}
We first ignore the quantization noise and evaluate the performance of the proposed two-stage algorithm to design the transformation matrices.
The empirical CDF curves of the average per-user rate achieved using $M=8$ and $M=32$ antennas per RRH are shown in Fig. \ref{fig:cdf_M8_ul} and \ref{fig:cdf_M32_ul}, respectively.
Firstly, it can be observed that the local CSI based DNN method, i.e., the first stage of the proposed two-stage DNN alone, can achieve better performance as compared to the EVD method in \cite{liu20171} and the modified EVD method in \cite{wiffen2021distributed}.
The performance gain comes from replacing the heuristic local loss function with the end-to-end loss function; it is obtained without introducing any additional communication costs.
This indicates that even if the RRHs do not have access to the realization of global CSI, the local DNNs can still learn to cooperate implicitly from the distribution of the global CSI through training.

Even though the local CSI based DNN can already outperform conventional local CSI based approaches, there is still a large gap from the global CSI based GD method.
For a system equipped with a small number of antennas, e.g., $M=8$, the dimension of the global CSI may not be very large; thus, it is feasible to send all the CSI matrices from the RRHs to the CP and use the GD method to design the transformation matrices based on the global CSI.
However, for large-scale antenna systems, e.g., $M=32$, the significant overhead requirement makes the global CSI based approach impractical.
In this case, the proposed iterative refinement scheme in the second stage can provide an efficient tradeoff between performance and communication overhead.
The convergence curve of the per-user rate with the number of signaling iterations is illustrated in Fig. \ref{fig:convergence_ul}, in which we provide a background bar plot that displays the ratio of communication overhead consumed by the proposed signaling scheme to that of the global CSI method.
It can be observed that the gap between the local CSI based DNN and the global CSI based GD method can be closed by $85\%$ with only $4$ refinement iterations, which corresponds to only $50\%$ of overhead required by the global CSI based approach.
Moreover, there is a remarkable boost in convergence speed of the meta-learning based GRU method as compared to the conventional GD approach, which comes from the use of the optimized update rule for this specific problem learned from training data.
This can significantly reduce overall communication rounds for CSI sharing and gradient transmission, leading to smaller communication overhead.

We further compare the convergence speed of GRU and conventional momentum based optimizers, including Adam\cite{kingma2014adam} and RMSProp\cite{Hinton2018Lecture6}.
It can be observed from Fig. \ref{fig:convergence_ul_adam} that across a range of step sizes for which Adam and RMSProp are tuned, their performances remain inferior to that of the trained GRU meta-learner. 
This performance gap can be attributed to the nature of Adam and RMSProp as general-purpose, hand-crafted optimizers, whereas the GRU has been specifically trained for this particular problem and has thus adapted to the input data distribution.

\subsubsection{Computational Complexity}

We compare the inference time complexity of the proposed scheme along with the benchmarks in Table \ref{tab:complexity}, where $d$ and  $\Tilde{d}$ represent the maximum dimensions of the hidden layers of the local CSI based DNN and the GRU, respectively. 
Since the number of neural network hidden layers ($d$ or $\Tilde{d}$) should scale proportionally with the problem size (e.g., $M$), the theoretical complexity scaling of the local CSI based DNN and the EVD method are seen as of comparable order. 
However, in practical implementations, matrix multiplication operations are highly optimized on the GPU hardware, while the EVD operation remains a computational bottleneck. 
We empirically validate this distinction using an Nvidia T4 GPU, as summarized in Table \ref{tab:complexity}, where it can be seen that the local CSI based DNN method runs much faster.

The complexity in the second refinement stage is mainly driven by the gradient computation, particularly due to the matrix inverse term $( \bar{\mathbf{F}}_n \bar{\mathbf{F}}_n^{\mathsf{H}} + \sigma_{\mathrm{ul}}^2\mathbf{I} )^{-1}$.
Once these gradients are computed and sent back to each RRH, the complexities for both the GRU based and the GD based updates are minimal.

\begin{table}[t]
\small
\centering
\caption{Time complexity and average computation time of each method.
Here, $d$ and $\Tilde{d}$ represent the maximum dimensions of hidden layers of local CSI based DNNs and GRU, respectively.}
% \color{blue}
\begin{tabular}{|l|l|l|}
\hline
Method  & Time Complexity &   \begin{tabular}[c]{@{}l@{}}Average\\ Time \end{tabular}  
\\ \hline
\begin{tabular}[c]{@{}l@{}}Local CSI\\ based DNN\end{tabular} & $\mathcal{O}(dMN + dKM + d^2)$ & 0.01s  
\\ \hline
\begin{tabular}[c]{@{}l@{}}Local CSI\\ based EVD\end{tabular}   & $\mathcal{O}(M^2N)$ & 0.37s \\ \hline
\begin{tabular}[c]{@{}l@{}}Gradient\\ computation \end{tabular} & $\mathcal{O} \left( B^2K^2N^2 + B^3K^3N \right)$  & 0.16 s 
\\ \hline
Each GRU iteration & $\mathcal{O}\left( \Tilde{d}KM  + \Tilde{d}^2 \right)$ & 0.00140 s \\ \hline
Each GD iteration    & $\mathcal{O}\left( KM \right)$ & 0.00003 s \\ \hline
\end{tabular}
\label{tab:complexity}
\end{table}

\subsubsection{Impact of Imperfect CSI}

\begin{figure}[t]
    \centering
    {\includegraphics[width=8cm]{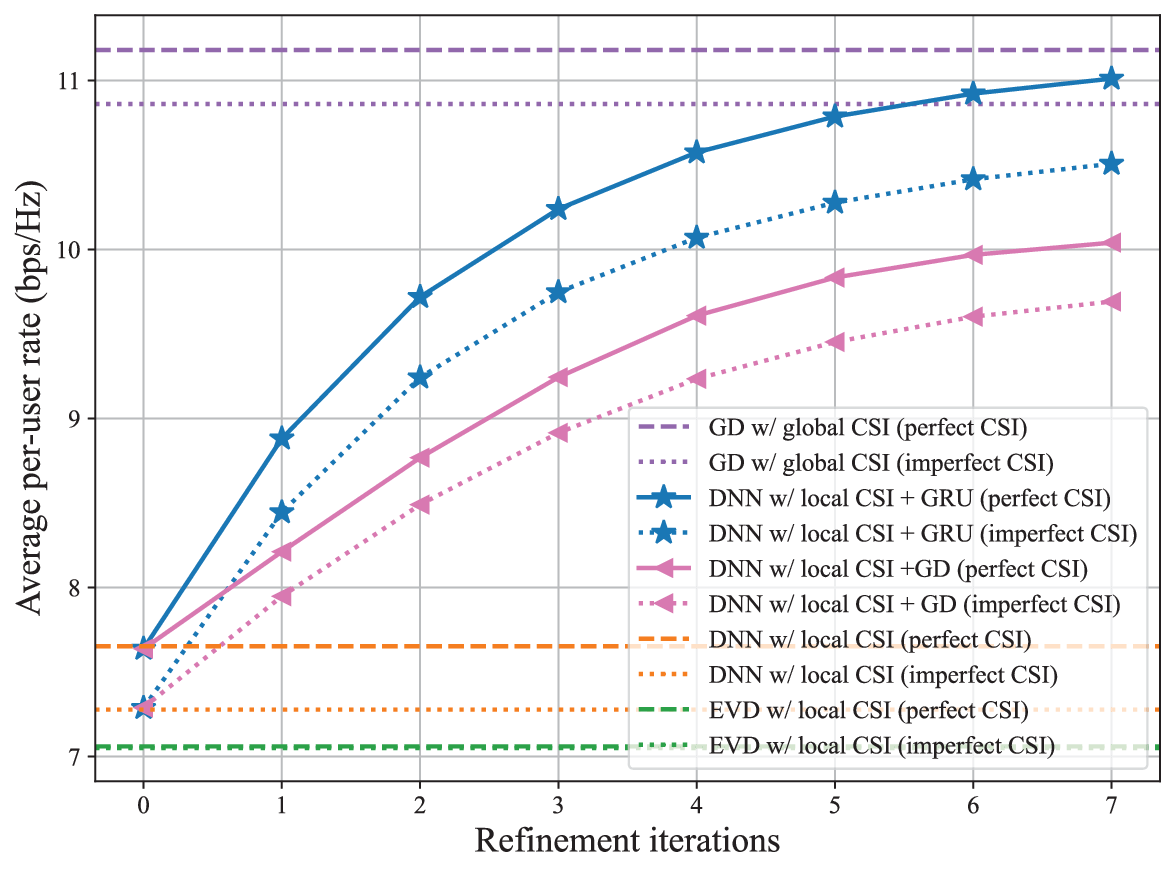}}
    % \vspace{-0.3cm}
    \caption{Comparison of uplink per-user rate assuming perfect and imperfect CSI ($M=32$).}
    % \vspace{-0.5cm}
    \label{fig:convergence_ul_CSI}
\end{figure}

Assuming that the channel estimation errors can be modeled as Gaussian random variables, we assess the sum rate performance under an estimation error variance of $-65$ dBm, as illustrated in Fig. \ref{fig:convergence_ul_CSI}.
It can be observed that the performances of all proposed approaches, as well as the global CSI based benchmark, degrade in the presence of imperfect CSI. 
However, the local CSI based DNN continues to outperform the EVD based approach, and the GRU still converges faster than GD. 
We also observe that the EVD method is more robust to channel estimation errors, but it has a large performance gap from the other approaches.

\begin{figure}[t]
    \centering
    {\includegraphics[width=8cm]{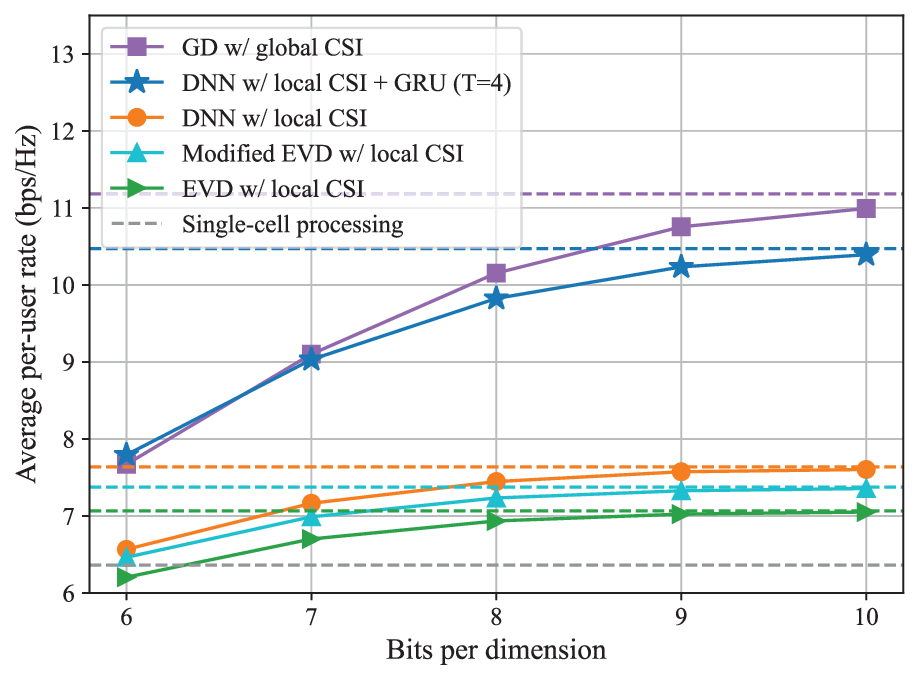}}
    % \vspace{-0.3cm}
    \caption{Uplink average per-user rate vs. the number of quantization bits per dimension using the uniform scalar quantizer ($M=32$).}
    % \vspace{-0.5cm}
    \label{fig:quant_ul}
\end{figure}

\subsubsection{Effects of Quantization}

With the optimized compression and expansion matrices designed using the proposed two-stage algorithm, we can evaluate the scalar quantization scheme in Section \ref{sec:quantization} in the optimized subspace.
The quantization scaling factor $\gamma$ is determined by manual line search.
As shown in Fig. \ref{fig:quant_ul}, the achievable per-user rate of each scheme considering quantization effects gradually converges to the rate without quantization with the increasing number of bits per fronthaul link.
Additionally, the proposed GRU method can significantly narrow the performance gap between local CSI and global CSI based methods for any number of bits per dimension.
Note that both the GRU method and the global CSI based benchmark require relatively more bits to converge to the rate without adding quantization.
This is due to their superior interference cancellation capabilities, which make them more susceptible to quantization noise in high-rate regimes.

\section{Downlink sum-rate maximization Problem with User-Centric Clustering}\label{sec:dl}

As the second example of the proposed algorithm, this section evaluates the performance of using the proposed framework to design the fronthaul compression strategy for the downlink channel model in \eqref{eq:rn_dl}.
Next, we introduce the downlink system model incorporating the user-centric clustering setup and present the simulation results.

\subsection{System Model}
Consider a user-centric clustering strategy, where each user is associated with its strongest RRHs. 
We use $\Theta _n$ to denote the serving cluster of RRHs for user $n$ and use $\left| \Theta _n \right|$ to denote the RRH cluster size. 
Further, from the perspective of each RRH, we use $\Phi_b$ to denote the associated users of RRH $b$, i.e., $\Phi_b=\left\{n: b \in \Theta_n, n=1, \dots, N\right\}$.

We use the two-stage algorithm in Section \ref{sec:design_W} to design the transformation matrix $\mathbf{W}_b^{\mathrm{dl}} \in \mathbb{C}^{K \times M}$, and represent the low-dimensional effective channel $\mathbf{F}_b^{\mathrm{dl}} \in \mathbb{C}^{K \times N}$ between RRH $b$ and all the users as in \eqref{eq:F_b_dl}.
Further, we use
$\mathbf{\bar{W}}_i^{\mathrm{dl}}\in \mathbb{C} ^{\left| \Theta _n \right|K\times \left| \Theta _n \right|M}$ to represent the block-diagonal transformation matrix for the RRH cluster $\Theta_i$, 
use $\bm{f}_{i,n}^{\mathrm{dl}}=\mathbf{\bar{W}}_{i}^{\mathrm{dl}} \left[ \cdots \left( \bm{h}_{bn}^{\mathrm{dl}} \right) ^{\mathsf{H}}\cdots \right] _{b\in \Theta _i}^{\mathsf{H}}  \in \mathbb{C} ^{\left| \Theta _i \right|K}$ to represent the effective channel vector between user $n$ and the RRHs in $\Theta_i$,
and use $\mathbf{\bar{F}}_i=\left[ \bm{f}_{i,1}^{\mathrm{dl}}\cdots \bm{f}_{i,N}^{\mathrm{dl}} \right] \in \mathbb{C} ^{\left| \Theta _i \right|K\times N}$ to represent the effective channel matrix between the RRHs in the cluster $\Theta_i$ and all the users.
With the effective CSI $\left\{ \mathbf{F}_b^{\mathrm{dl}} \right\}_{b=1}^B$ from all the RRHs, 
the transmit beamformer for user $n$, denoted by  
${\bm{c}}_{n}^{\mathrm{dl}}=\left[ \cdots \left( {\bm{c}}_{bn}^{\mathrm{dl}} \right) ^{\mathsf{H}}\cdots \right] _{b\in \Theta _n}^{\mathsf{H}}\in \mathbb{C} ^{\left| \Theta _n \right|K}$ can now be designed using the effective channel matrix 
between the RRHs in its cluster $\Theta_i$ and all the users (while ignoring the effect of quantization) as in \eqref{eq:c_n_dl_sec2_0}, where $\mathbf{F}^{\mathrm{dl}}$ and $\bm{f}_{n}^{\mathrm{dl}}$ are replaced by $\mathbf{\bar{F}}_n^{\mathrm{dl}}$ and $\bm{f}_{n,n}^{\mathrm{dl}}$, respectively.

With fixed beamformers $\left\{c_n^{\mathrm{dl}} \right\}_{n=1}^N$, in the data transmission stage the CP computes the beamformed signal vector $\bm{v}_b^{\mathrm{dl}} \in \mathbb{C}^K$ intended for RRH $b$ as:
\begin{equation}
   \bm{v}_{b}^{\mathrm{dl}}=\sum_{n\in \Phi _b}{\sqrt{p_n^{\mathrm{dl}}}\bm{{c}}_{bn}^{\mathrm{dl}}s_{n}^{\mathrm{dl}}} ,
\end{equation}
where the power factors $\left\{ p_n^{\mathrm{dl}} \right\}_{n=1}^N$ are designed to account for the per RRH power constraints \cite{dai2016energy, pan2017joint, park2013joint} or possibly per antenna power constraints \cite{kim2013low}.
For simplicity, this paper assumes a sum downlink transmit power constraint $P^{\mathrm{dl}}$ across all RRHs and equally assigns it to each user's beamforming vector across its serving RRHs, i.e., set $p_{n}^{\mathrm{dl}}=P^{\mathrm{dl}}/\mathrm{N}$.

We remark that incorporating per-RRH power constraints into our framework is feasible, but it requires more complex downlink beamformer designs such as the WMMSE algorithm\cite{chowdhury2021unfolding, shi2011iteratively}.
Additionally, within a user-centric clustering framework, the clustering scheme and power allocation need to be jointly designed, as outlined in \cite{dai2014sparse}.
Nonetheless, in the high SNR regime where CRAN operates, transmit MMSE beamforming with equal power allocation across the beams performs closely to the optimal beamformer\cite{bjornson2014optimal}.

Next, the signal $\bm{v}_b^{\mathrm{dl}}$ is quantized into 
$\bm{\hat{v}}_{b}^{\mathrm{dl}}$ using a $Q_b$-bit uniform scalar quantizer as in Section \ref{sec:quantization}, and
then forwarded onto the downlink fronthaul.
At RRH $b$, the transformation matrix $\mathbf{W}_{b}^{\mathrm{ul}}\in \mathbb{C} ^{K\times M}$ expands $\hat{\bm{v}}_{b}^{\mathrm{dl}} \in \mathbb{C} ^{K}$ to the signal $\bm{x}_{b}^{\mathrm{dl}} \in \mathbb{C} ^{M}$, i.e., $\bm{x}_{b}^{\mathrm{dl}}=\left( \mathbf{W}_{b}^{\mathrm{ul}} \right) ^{\mathsf{H}}\hat{\bm{v}}_{b}^{\mathrm{dl}}$.
{Each user $n$ receives} the superposition of signals transmitted from all the RRHs over the air as:
\begin{equation}
    \begin{aligned}
    y_{n}^{\mathrm{dl}}&=\sqrt{p_{n}^{\mathrm{dl}}}\left( \bm{f}_{n,n}^{\mathrm{dl}} \right) ^{\mathsf{H}}\bm{c}_{n}^{\mathrm{dl}}s_{n}^{\mathrm{dl}}
    \\
    &+\sum_{i\ne n}{\sqrt{p_{i}^{\mathrm{dl}}}\left( \bm{f}_{i,n}^{\mathrm{dl}} \right) ^{\mathsf{H}}\bm{c}_{i}^{\mathrm{dl}}s_{i}^{\mathrm{dl}}}+\left( \bm{f}_{n}^{\mathrm{dl}} \right) ^{\mathsf{H}}\bm{e}+z_{n}^{\mathrm{dl}},
\end{aligned}
\end{equation}
where the vector $\bm{e}^{\mathrm{dl}} \sim \mathcal{CN}\left(0,\mathbf{{D}}^{\mathrm{dl}} \right)$ is the  quantization noise vector with covariance $\mathbf{D}^{\mathrm{dl}}=\mathrm{diag}\left( {{D}}_{11}^{\mathrm{dl}},\cdots,{{D}}_{BK}^{\mathrm{dl}}  \right) \in \mathbb{C}^{BK \times BK}$ across all the RRHs.
{Finally}, the downlink achievable rate of user $n$ can be evaluated as:
\begin{equation}
\begin{aligned}
        &\tilde{R}_{n}^{\mathrm{dl}}=
        \\ 
        &\log \left(1+ \frac{\left| \left( \bm{c}_{n}^{\mathrm{dl}} \right) ^{\mathsf{H}}\bm{f}_{n,n}^{\mathrm{dl}} \right|^2}{\sum \limits_{\substack{ i\ne n}}{\left| \left( \bm{c}_{i}^{\mathrm{dl}} \right) ^{\mathsf{H}}\bm{f}_{i,n}^{\mathrm{dl}} \right|^2}+\left| \left( \bm{f}_{n}^{\mathrm{dl}} \right) ^{\mathsf{H}}\mathbf{D}^{\mathrm{dl}}\bm{f}_{n}^{\mathrm{dl}} \right|^2+\sigma _{\mathrm{dl}}^{2}} \right) .
\end{aligned}
\label{eq:rate_R_ori_dl}
\end{equation}
Note that the rate $R_n^{\mathrm{dl}}$ ignoring the quantization error can be expressed by excluding the $| ( \bm{f}_{n}^{\mathrm{dl}} ) ^{\mathsf{H}}\mathbf{D}^{\mathrm{dl}}\bm{f}_{n}^{\mathrm{dl}} |^2$ term from \eqref{eq:rate_R_ori_dl}.

\subsection{Numerical Results}

\begin{figure}[t]
    \centering
    {\includegraphics[width=8cm]{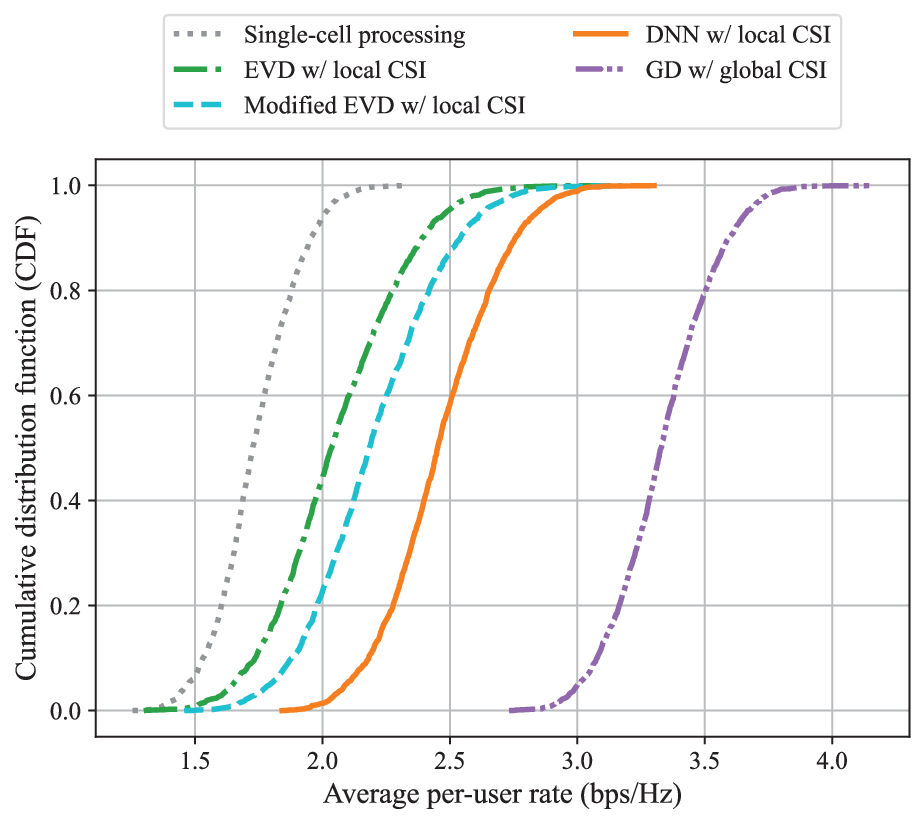}}
    \caption{Empirical CDF curve of downlink average per-user rate ($M=8$).}
    \label{fig:cdf_M8_dl}
\end{figure}

\begin{figure}[t]
    \centering
    {\includegraphics[width=8cm]{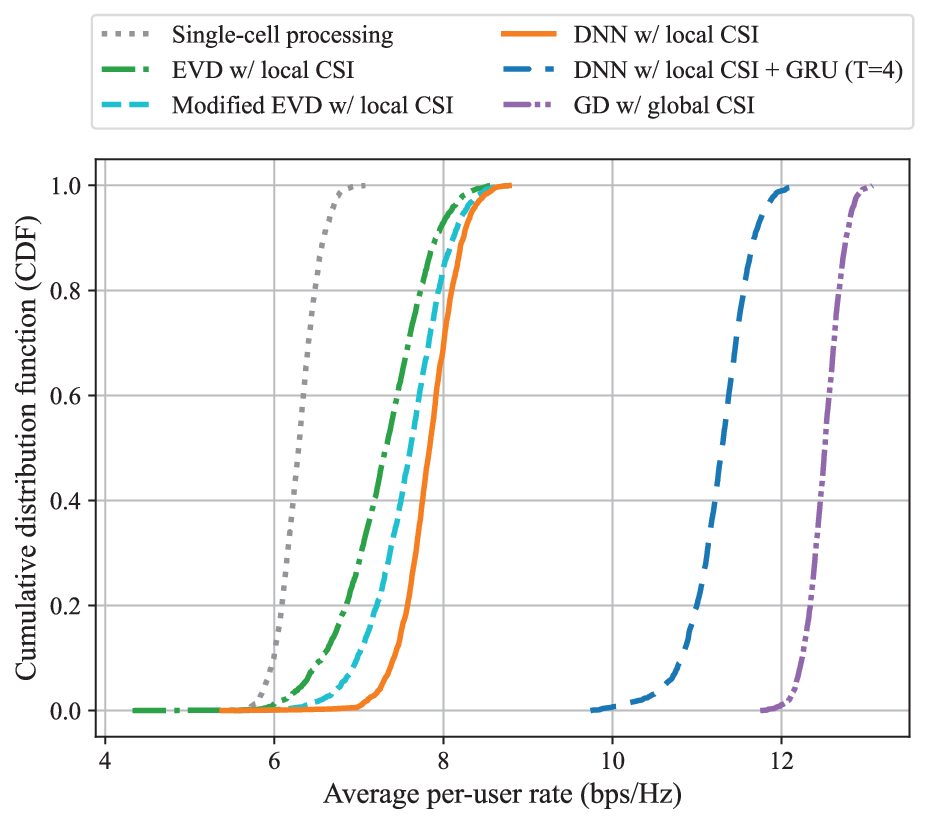}}
    \caption{Empirical CDF curve of downlink average per-user rate ($M=32$).}
    \label{fig:cdf_M32_dl}
\end{figure}

\begin{figure}[t]
    \centering
    {\includegraphics[width=8cm]{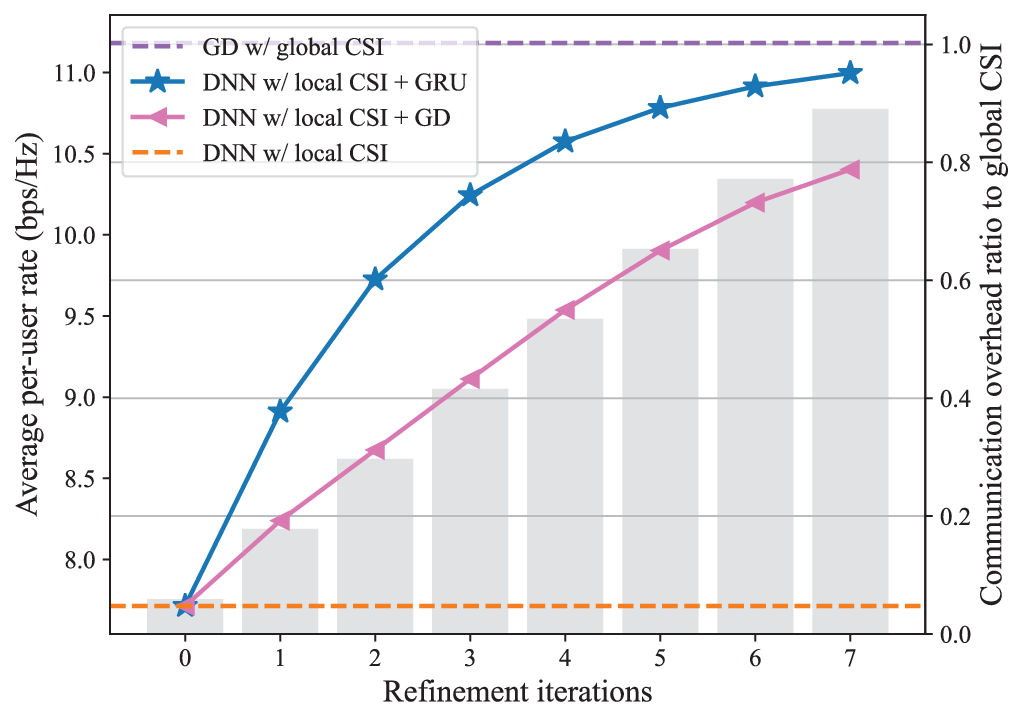}}
    \caption{Left vertical axis: convergence curve for the downlink per-user rate. Right vertical axis: bar graph indicating the communication overhead ratio of the proposed iterative scheme to the global CSI based method ($M=32$).}
    \label{fig:convergence_dl}
    % \vspace{-0.3cm}
\end{figure}

\begin{figure}[t]
    \centering
    {\includegraphics[width=8cm]{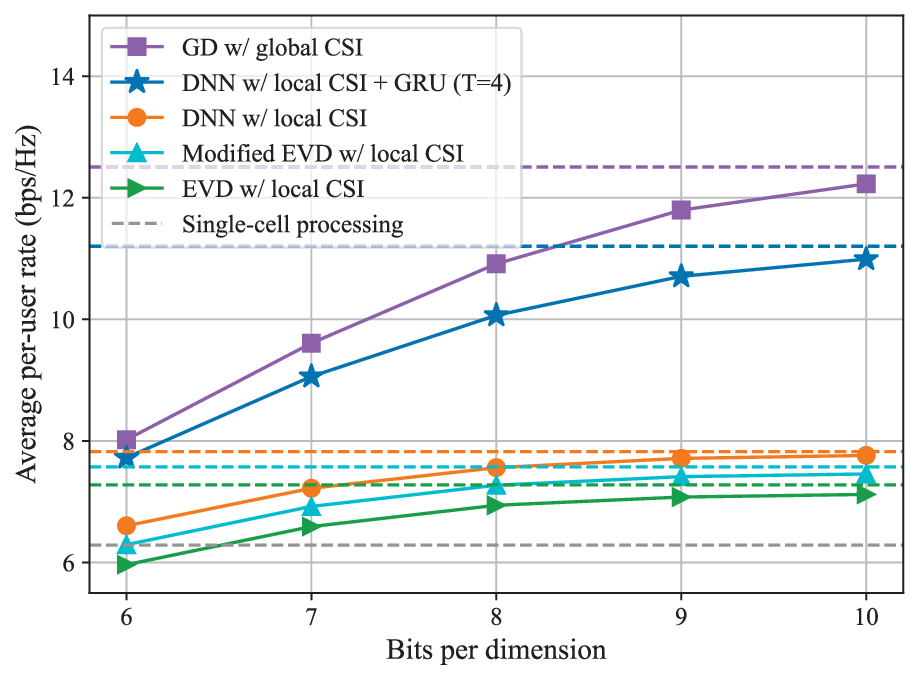}}
    \caption{Downlink average per-user rate vs. the number of quantization bits per dimension using the uniform scalar quantizer ($M=32$).}
    \label{fig:quant_dl}
\end{figure}

\subsubsection{Simulation Setup}
We set the total downlink transmit power budget across the 19 RRHs to $P^{\mathrm{dl}}=42.8$dBm.
Other simulation configurations, including simulation topology, parameters, and DNN training configurations, are identical to the uplink simulations as described in Section \ref{sec:sim_ul}.
The performance of the proposed data-driven approach is compared against the same benchmark schemes outlined in Section \ref{sec:sim_ul}.

\subsubsection{Performance of the Proposed Two-Stage Algorithm}

We apply the proposed two-stage framework to design the downlink dimension-expansion matrices.
Firstly, Fig. \ref{fig:cdf_M8_dl} and Fig. \ref{fig:cdf_M32_dl} show that the local CSI based DNN method has an advantage over the EVD and modified EVD method.
Moreover, Fig. \ref{fig:convergence_dl} shows that the performance of the GRU method can converge to that of the global CSI based benchmark.
For example, the gap between the local CSI and global CSI based benchmark can be closed by $75\%$ using $50\%$ of overhead required by the global CSI based approach.
However, the convergence rate in the downlink is relatively slower as compared to that in the uplink. 
This is because the GRU needs more iterations to converge when the signal-to-noise ratio is higher.
Nonetheless, the convergence rate of the GRU method is still much faster than that of GD.

\subsubsection{Effects of Quantization}

We examine the performance of the downlink system sum rate when the quantization distortion effects are considered.
The achievable rate after quantization can gradually converge to the rate without considering quantization noise as the number of quantization bits increases. 
Moreover, the GRU method can considerably narrow the performance gap between the local CSI and global CSI based methods for a specified number of quantization bits.

\section{Conclusion}\label{sec:conclusion}

This paper investigates the problem of designing fronthaul compression schemes including linear transformation and uniform scalar quantization in a C-RAN system to maximize the system sum rate.
A two-stage deep learning framework is proposed to design the transformation matrices, which are derived from the local CSI using fully connected neural networks in the first stage and further refined iteratively using the downlink signaling from the CP in the second stage.
To reduce the communication overhead, a low-dimensional signaling scheme is proposed to reduce the overhead for each iteration, and a novel GRU based meta-learning framework is proposed to accelerate the convergence speed of the refinement process.
Further, uniform scalar quantizers are designed to compress the dimension-reduced signals.
The performance of the proposed algorithm is evaluated in the uplink and downlink under a user-centric clustering setting.
For both applications, simulation results show that the performance of the proposed neural network can quickly converge to that of the global CSI based benchmark with significantly smaller overhead and that using the first stage alone can already outperform the local CSI based EVD benchmark.

%%%%%%%%%%%%%%%%%%%%%%%%%%%%%%

\bibliographystyle{IEEEtran} % We choose the "plain" reference style
\bibliography{reference} % Entries are in the refs.bib file

\vfill

\end{document}